\documentclass[a4paper,11pt]{article}

\usepackage{setup}
\usepackage{jheppub}
\usepackage{todonotes}
\usepackage{placeins}
\usepackage{longtable}

\usepackage{cite}
\usepackage{ifthen}
\usepackage{subcaption}
\usepackage{siunitx}
\usepackage{enumitem}
\usepackage{booktabs}
\usepackage{bookmark}

\usepackage{cleveref}
\DeclareSIUnit\fb{\femto\barn}

\def\lapprox{\lower .7ex\hbox{$\;\stackrel{\textstyle <}{\sim}\;$}}
\def\gapprox{\lower .7ex\hbox{$\;\stackrel{\textstyle >}{\sim}\;$}}
\definecolor{lightgray}{HTML}{A6A39A}
\definecolor{darkgray}{HTML}{504E48}
\definecolor{silver}{HTML}{E0DFDE}
\definecolor{brown}{HTML}{5F4541}
\definecolor{beige}{HTML}{DCCCAC}
\definecolor{green}{HTML}{345F53}
\definecolor{yellow}{HTML}{F6B65A}
\definecolor{blue}{HTML}{568BCF}
\definecolor{red}{HTML}{AE1932}
\definecolor{orange}{HTML}{D16F15}
\definecolor{darkpurple}{HTML}{39FF14}

\newcommand{\expo}[1]{\ensuremath{\mathrm{e}^{#1}}}
\newcommand{\calF}{\ensuremath{\mathcal{F}}}

\makeatletter
\newcommand{\myitem}[1]{%
	\item[#1]\protected@edef\@currentlabel{#1}%
}
\makeatother

\preprint{{\raggedleft%
IPPP/25/64, ZU-TH 67/25, MCNET-25-25 \\
}}

\title{The Thrust Distribution at NNLO+NNLL in Higgs Decays to Quarks and Gluons}
\author[a]{Elliot~Fox,}
\author[b,c]{Aude~Gehrmann--De Ridder,}
\author[c]{Thomas~Gehrmann,}
\author[a,d]{Nigel~Glover,}
\author[a]{Matteo~Marcoli,}
\author[e]{Christian~T.~Preuss}

\affiliation[a]{Institute for Particle Physics Phenomenology, Department of Physics, University of Durham, Durham, DH1 3LE, UK}
\affiliation[b]{Institute for Theoretical Physics, ETH, 8093 Z{\"u}rich, Switzerland}
\affiliation[c]{Physik-Institut, Universit{\"a}t Z{\"u}rich, 8057 Z{\"u}rich, Switzerland}
\affiliation[d]{CERN, 1211 Geneva 23, Switzerland}
\affiliation[e]{Institut f{\"u}r Theoretische Physik, Georg-August-Universit{\"a}t G{\"o}ttingen, 37077 G{\"o}ttingen, Germany}

\emailAdd{elliot.fox@durham.ac.uk}
\emailAdd{thomas.gehrmann@uzh.ch}
\emailAdd{gehra@phys.ethz.ch}
\emailAdd{e.w.n.glover@durham.ac.uk}
\emailAdd{matteo.marcoli@durham.ac.uk}
\emailAdd{christian.preuss@uni-goettingen.de}

\abstract{
  We present a calculation of the thrust distribution in Higgs decays to quarks and gluons, $H\to b\bar{b}$, $H\to c\bar{c}$, and $H\to gg$, including the resummation of large logarithmic corrections that arise in the two-particle limit at next-to-next-to-leading logarithmic (NNLL) accuracy, and match it to fixed-order results for three-particle decays at next-to-next-to-leading order (NNLO) in the strong coupling.
  The resummation is performed analytically within the \ARES framework and combined with the fixed-order results using the logR matching technique.
  The fixed-order calculation is carried out numerically with the \NNLOJET parton-level event generator, using the antenna subtraction method.
  We perform detailed cross-validation in the two-particle region, demonstrating that the expansion of the NNLL resummed result correctly reproduces the logarithmic structure of the fixed-order calculation to $\order{\alphas^3}$, up to a predictable \N{3}LL term at $\order{\alphas^3L}$.
  In addition to providing the first NNLO+NNLL accurate predictions for the thrust distribution in Higgs decays to quarks and gluons, we analytically extract the $\order{\alphas^2}$ hard-virtual correction $c_2$ and the $\alphas^3L$ term $G_{31}$ in both the $H\to q\bar{q}$ ($q=b,c$) and $H\to gg$ decay channels.
}

\begin{document}
\maketitle
\flushbottom

\section{Introduction}
\label{sec:intro}
The thrust event-shape observable $T$ \cite{Brandt:1964sa,Farhi:1977sg} measures the degree of isotropy of a scattering event and represents one of the most extensively studied observables in both experimental and theoretical particle physics.
In most analyses, the complementary variable $\tau=1-T$ is employed, as it conforms to the standard convention for event shapes, approaching zero in the two-particle limit.
Its definition in terms of final-state momenta is given by
\begin{equation}
  \tau = \min\limits_{\vec{n}}\left(1-\frac{\sum\limits_i \mods{\vec{p}_i\cdot\vec{n}}}{\sum\limits_i \mods{\vec{p}_i}} \right) \,.
\end{equation}
According to this definition, a pencil-like configuration, with two back-to-back jets, corresponds to $\tau = 0$, while the limit $\tau \to \frac{1}{2}$ describes an isotropic event.
To have a non-vanishing value of $\tau$, it is necessary to consider events with at least three particles in the final state.
For events containing exactly three particles, it holds that $T \geq \frac{2}{3}$ and $\tau \leq \frac{1}{3}$.

Fixed-order calculations of the thrust observable in electron-positron annihilation, $e^+e^-\to \text{jets}$, have a long history, beginning with the pioneering next-to-leading order (NLO) calculation nearly half a century ago~\cite{Ellis:1980wv}.
Next-to-next-to-leading order (NNLO) corrections were first obtained in Ref.~\cite{Gehrmann-DeRidder:2007nzq} and have been thoroughly validated~\cite{Weinzierl:2009ms,DelDuca:2016csb,DelDuca:2016ily}.

Fixed-order predictions of event shapes, such as thrust, are only reliable for resolved three-parton configurations, i.e., sufficiently far away from the $\tau = 0$ region.
As $\tau$ becomes small, the perturbative expansion in the strong coupling $\alphas$ is spoiled by large logarithmic contributions in $\log(\tau)$, which need to be resummed.

In the two-particle limit, the thrust event shape can be expressed in terms of the transverse momentum $k_\mathrm{t}^{(\ell)}$ and rapidity $\eta^{(\ell)}$ of a soft-collinear emission with respect to leg $\ell$ as
\begin{equation}
  \tau \sim \sum\limits_\ell \frac{k_\text{t}^{(\ell)}}{Q}\expo{-\eta^{(\ell)}}\,,
  \label{eq:thrustParam}
\end{equation}
where the sum runs over partonic legs $\ell$.
In this form, the additive nature of the thrust observable is made manifest, i.e. its value can be determined from the sum of the contributions of individual radiating legs.
This admits a relatively simple calculation of large logarithmic enhancements in the small-$\tau$ limit, which has been exploited to compute the next-to-leading logarithmic (NLL) correction for thrust in quark-antiquark final states in Laplace space in \cite{Catani:1991kz}.
Its extension to the next-to-next-to-leading logarithmic (NNLL) order has been performed in \cite{Monni:2011gb}, after an equivalent calculation has been carried out in the framework of Soft-Collinear Effective Theory (SCET) \cite{Becher:2008cf}.
Higher-logarithmic corrections to the thrust distribution in electron-positron annihilation have been obtained at next-to-next-to-next-to-leading logarithmic (\N{3}LL) \cite{Becher:2008cf,Abbate:2010xh} and up to next-to-next-to-next-to-next-to-leading logarithmic (\N{4}LL) order \cite{Aglietti:2025jdj}.

The availability of high-precision theory predictions for the thrust has enabled a multitude of studies to extract the value of the strong coupling constant from experimental LEP data, see e.g.~\cite{OPAL:2004wof,L3:2004cdh,Becher:2008cf,Bethke:2009ehn,Abbate:2010xh,Dissertori:2009qa,Dissertori:2009ik,OPAL:2011aa,Benitez:2024nav,Farren-Colloty:2025amh,Nason:2025qbx}. 
In this context, non-perturbative power corrections are important to accurately describe the experimental data.
These were first discussed \cite{Webber:1994cp} and calculated \cite{Dokshitzer:1995zt} about thirty years ago, but still remain an active topic of research to date \cite{Abbate:2010xh,Agarwal:2020uxi,Caola:2021kzt,Bhattacharya:2022dtm,Caola:2022vea,Nason:2023asn,Dasgupta:2024znl,Hoang:2025uaa,Aglietti:2025jdj}.

With future lepton colliders~\cite{FCC:2018byv,FCC:2018evy,CEPCStudyGroup:2018ghi,ILC:2013jhg} projected to operate as ``Higgs factories'', where Higgs bosons are copiously produced, interest in precise predictions of event shapes in hadronic Higgs decays has increased~\cite{Gao:2016jcm,Gao:2019mlt,Gao:2020vyx,Knobbe:2023njd,Coloretti:2022jcl,Gehrmann-DeRidder:2023uld,Gehrmann-DeRidder:2024avt,Fox:2025cuz,Fox:2025qmp}. 
In contrast to the situation at LEP, event-shape observables in hadronic Higgs decays probe both the quark-antiquark and the gluon-gluon final states, since the most dominant hadronic Higgs decay channels are $H\to b\bar{b}$ and $H\to gg$.
Decays through intermediate $WW^*$, $ZZ^*$ and  $\tau\tau$ states are assumed to be separable based on their characteristic kinematical signatures~\cite{Ma:2024qoa}.
Building on earlier NLO results for $H\to gg$~\cite{Gao:2019mlt,Coloretti:2022jcl} and NNLO calculations of closely related jet observables in $H\to b\bar{b}$~\cite{Mondini:2019gid,Mondini:2019vub}, the full NNLO QCD corrections to the thrust distribution
in $H\to b\bar{b}$, $H\to c\bar{c}$, and $H\to gg$ decays has been obtained recently in \cite{Fox:2025qmp}, using the implementation of Higgs decays to quarks and gluons~\cite{Fox:2025cuz} in \NNLOJET \cite{NNLOJET:2025rno}.
Resummed predictions of thrust in Higgs decays have first been achieved at NNLL in \cite{Mo:2017gzp} using SCET and later extended to \N{3}LL and \N{4}LL in \cite{Ju:2023dfa}.
Matched predictions of thrust in Higgs decays to quarks and gluons have been performed in \cite{Gehrmann-DeRidder:2024avt} at NLO+NLL$^\prime$.

In this work, we present the NNLL resummation of the thrust distribution in Higgs decays to quarks and gluons, matched to fixed-order results at NNLO. The resummation is performed analytically within the \ARES framework~\cite{Banfi:2014sua,Banfi:2018mcq}, while the fixed-order calculation is carried out numerically in the \nnlojet framework \cite{NNLOJET:2025rno}.

The structure of the paper is as follows.
Sec.~\ref{sec:framework} outlines the theoretical framework of the calculation, including details on the NNLO fixed-order computation, the NNLL resummation, and the matching procedure.
Sec.~\ref{sec:results} describes the practical setup of the calculation and presents the main results, namely the $\text{NNLO}+\text{NNLL}$ accurate predictions for the thrust distribution in $H\to gg$, $H\to b\bar{b}$, and $H\to c\bar{c}$ decays.
Section~\ref{sec:conclusion} concludes this work and provides an outlook on future developments.

\begin{figure}[t]
  \centering
  \includegraphics[width=0.9\textwidth]{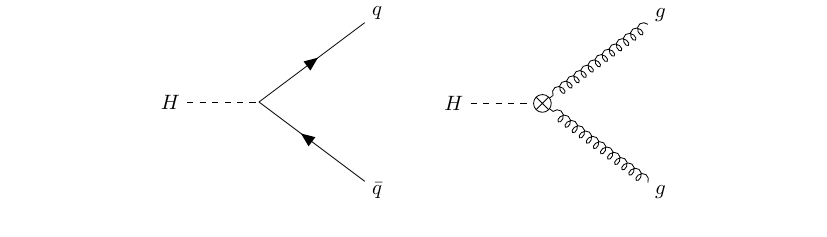}
  \caption{Hadronic Higgs decay categories: $H\to q\bar{q}$ with a Yukawa coupling  (left) and $H\to gg$ via an effective coupling (right).}
  \label{fig:diagH2jLO}
\end{figure}

\section{Setup of the Calculation}
\label{sec:framework}
Our calculation is based on an effective field theory in which the Higgs boson couples to gluons via an effective $Hgg$ coupling~\cite{Wilczek:1977zn,Shifman:1978zn,Inami:1982xt} and to kinematically massless quarks via a non-vanishing Yukawa coupling, cf.\ Fig.~\ref{fig:diagH2jLO}.
In this setup, the LO inclusive hadronic decay widths are given as
\begin{equation}
  \Gamma_{H\to q\bar{q}}^{(0)} = \frac{y_q^2(\muR)m_H\NC}{8\pi} \,, \quad \Gamma_{H\to gg}^{(0)} = \frac{\lambda_0^2(\muR)m_H^3(\NC^2-1)}{64\pi} \, ,
  \label{eq:ratesLO}
\end{equation}
where $m_H$ denotes the Higgs mass.
The Yukawa coupling, $y_q$, and the LO effective $Hgg$ coupling in the heavy-top limit, $\lambda_0$, determine the couplings of quarks and gluons to the Higgs boson.
They are given by
\begin{equation}\label{eq:couplings}
  y_q^2(\muR) = m_q^2(\muR)\sqrt{2}\GF\,, \quad \lambda_0^2(\muR) = \frac{\alphas^2(\muR)\sqrt{2}\GF}{9\pi^2}\,,
\end{equation}
in terms of the Fermi constant $\GF$.
Both the Yukawa and the effective Higgs-gluon coupling are renormalised in the $\overline{\text{MS}}$ scheme and we evaluate these quantities at $\muR$ with $\NF=5$. 
The running of the quark masses is taken into account according to \cite{Vermaseren:1997fq}. As illustrated in the following, our results are typically normalised with respect to inclusive decay widths. For this reason, when presenting numerical results for the $H\to q\bar{q}$ channel, we do not specify the flavour of the quarks coupling to the Higgs boson via Yukawa interaction, as our results are valid for any flavour. In Sec.~\ref{sec:results_matched}, where we consider the sum over all decay channels, we suitably indicate $q\bar{q}=b\bar{b},\,c\bar{c}$.

\subsection{Fixed-Order Calculation at NNLO}
\label{subsec:nnlo}
Up to NNLO in fixed-order perturbation theory, the three-particle decay rate of the Higgs boson in the decay channel $X$, can be written differentially in thrust as
\begin{equation}
  \begin{split}
    \frac{1}{\Gamma^{(0)}_{H\to X}(m_H,\mu_R)}\frac{\rd \Gamma_{H\to X}(m_H,\mu_R)}{\rd \tau} &= \\
    &\hspace{-5cm}\left(\frac{\alphas(\mu_R)}{2\pi}\right)\frac{\rd A_X(\mu_R)}{\rd \tau} + \left(\frac{\alphas(\mu_R)}{2\pi}\right)^2\frac{\rd B_X(\mu_R)}{\rd \tau} + \left(\frac{\alphas(\mu_R)}{2\pi}\right)^3\frac{\rd C_X(\mu_R)}{\rd \tau} + \mathcal{O}(\alpha_s^4)\,.
  \end{split}
  \label{eq:rateDiff}
\end{equation}
Here, the decay rate is normalised to the respective Born-level $H\to X$ decay width $\Gamma^{(0)}_{H\to X}$, with $X=q\bar{q}$, $gg$, and $\muR$ denotes the renormalisation scale.
The coefficients $A_X$, $B_X$, and $C_X$ are the dimensionless LO, NLO, and NNLO coefficients, respectively.
The LO coefficient $A_X$ is finite for $\tau > \tau_0$, where $\tau_0 > 0$ denotes a suitable lower cutoff on the thrust observable.
To calculate the NLO and NNLO coefficients $B_X$ and $C_X$, a suitable subtraction scheme must be employed to remove explicit infrared poles in virtual amplitudes and implicit infrared divergences in real-radiation matrix elements.
Technically speaking, the calculation of $A_X$, $B_X$ and $C_X$ for thrust closely resembles the calculation of 
the respective contributions to the three-jet decay rate.

We calculate the perturbative coefficients $A_X$, $B_X$, and $C_X$ using the \NNLOJET parton-level Monte-Carlo event generator \cite{NNLOJET:2025rno}, which employs the antenna-subtraction scheme \cite{Gehrmann-DeRidder:2005btv,Currie:2013vh}.
Recently, \NNLOJET has been used to calculate NNLO-accurate predictions for a wide range of event-shape observables \cite{Fox:2025qmp} in hadronic Higgs decays to three jets and \N{3}LO predictions for two-jet rates \cite{Fox:2025cuz} in the $H\to b\bar{b}$, $H\to c\bar{c}$, and $H\to gg$ channels.
The Higgs decay processes to quarks and gluons in \NNLOJET make use of subtraction terms constructed from generalised antenna functions \cite{Fox:2024bfp}, which are derived directly from the relevant infrared limits using the algorithm described in \cite{Braun-White:2023sgd,Braun-White:2023zwd}.

At NLO and NNLO, we normalise \eqref{eq:rateDiff} by the inclusive NLO or NNLO decay width, $\Gamma^{(1)}_{H\to X}$ or $\Gamma^{(2)}_{H\to X}$, respectively.
In terms of their LO results~\eqref{eq:ratesLO}, the higher-order rates can be expressed as
\begin{equation}
  \Gamma^{(k)}_{H\to X} = \Gamma^{(0)}_{H\to X}\, \left(1+\sum\limits_{n=1}^k\alphas^n H_{X}^{(n)}\right)\,,
  \label{eq:ratesNkLO}
\end{equation}
where the relevant corrections $H_{X}^{(n)}$ up to NNLO are given e.g.~in \cite{Herzog:2017dtz}.
We summarise them in Appendix~\ref{app:A}.
Expanding the normalisation \eqref{eq:ratesNkLO}, we obtain the NNLO differential decay rate
\begin{equation}
  \begin{split}
    \frac{1}{\Gamma^{(2)}_{H\to X}(m_H,\mu_R)}\frac{\rd \Gamma_{H\to X}(m_H,\mu_R)}{\rd \tau} &= \\
    &\hspace{-5cm} \left(\frac{\alphas(\mu_R)}{2\pi}\right)\frac{\rd \bar{A}_X(\mu_R)}{\rd \tau} + \left(\frac{\alphas(\mu_R)}{2\pi}\right)^2\frac{\rd \bar{B}_X(\mu_R)}{\rd \tau} + \left(\frac{\alphas(\mu_R)}{2\pi}\right)^3\frac{\rd \bar{C}_X(\mu_R)}{\rd \tau} + \mathcal{O}(\alpha_s^4) \,.
  \end{split}
  \label{eq:rateDiffNorm}
\end{equation}
The coefficients $\bar{A}_X$, $\bar{B}_X$, and $\bar{C}_X$ are then related to $A_X$, $B_X$, and $C_X$ in \eqref{eq:ratesNkLO} by
\begin{equation}
  \bar{A}_X = A_X\,, \quad \bar{B}_X = B_X - H_{X}^{(1)}A_X\,, \quad \bar{C}_X = C_X - H_{X}^{(1)}B_X + \left(\left(H_{X}^{(1)}\right)^2-H_{X}^{(2)}\right)A_X \,.
\end{equation}
We illustrate the renormalisation-scale dependence of the perturbative expansion coefficients in App.~\ref{app:B}.

\subsection{Resummation at NNLL}
\label{subsec:nnll}
The resummation of large logarithmic corrections is performed using the 
cumulative cross section
\begin{align}
  \Sigma_{H\to X}(\tau) &= \frac{1}{\Gamma_{H\to X}}\int\limits^{\tau}_0\, \frac{\rd \Gamma_{H\to X}}{\rd \tau'}\, \rd\tau' \nonumber\\
  &= \left(1+\sum\limits_n\left(\frac{\alphas}{2\pi}\right)^n c_{n}^{(H\to X)}\right)\expo{-R_X(\lambda)}\calF_X(R_X^\prime(\lambda))\,,
  \label{eq:cumulant}
\end{align}
where the Sudakov radiator $R_X$ accounts for single-emission effects, while $\calF_X$ encodes the effect of multiple emissions on the parton ensemble.
It is customary to write the above expression in terms of the large logarithm $L=-\log(\tau)$ as $\lambda = \alphas\beta_0L$, where $\beta_0$ is the first-order coefficient in the QCD $\beta$-function (see App.~\ref{app:A} for our conventions).
The process-dependent coefficients $c^{(H\to X)}_{n}$ account for the non-logarithmic, constant contributions in the soft limit and induce an overall normalisation in the cumulative cross section.
We follow \cite{Banfi:2004yd} and define the logarithmic accuracy through the terms appearing in the exponent, as opposed to its expansion.

We implement the resummation of the thrust event-shape observable up to NNLL in a stand-alone code, which we interface to the output of \NNLOJET.
Our implementation is based on the \ARES scheme \cite{Banfi:2014sua,Banfi:2018mcq}.
Within the \ARES scheme, a recursively infrared-safe observable $V$ can be parametrised in terms of the transverse momentum $k_\text{t}^{(\ell)}$, rapidity $\eta^{(\ell)}$, and azimuthal angle $\phi$ for a single soft-collinear emission with momentum $k$ off leg $\ell$ as
\begin{equation}
  V_\ell(\{\tilde{p}\}, k) = d_\ell g_\ell(\phi) \left(\frac{k_\text{t}^{\ell}}{\mu_Q}\right)^a\expo{-b_\ell \eta^{(\ell)}} \,.
\end{equation}
Comparing this to \eqref{eq:thrustParam} yields the parameters $a=1$, $b_\ell \equiv b = 1$, $d_\ell=1$, and $g_\ell(\phi)=1$ for thrust.

In the following, we will explicitly review the necessary ingredients to the Sudakov radiator $R_X$ and multiple-emission function $\calF_X$ to resum the thrust observable $\tau$ in the \ARES scheme up to NNLL for both quark and gluon radiators. These quantities depend on the process through the identity of the radiating legs.  In our case $H \to X$, we are interested in processes with either two gluons ($X = gg$) or a quark-antiquark pair ($X = q\bar{q}$).  To streamline the notation, for the remainder of this section, we drop the process label, $X$ or $H \to X$, in favour of the more usual identity of the radiating legs $\ell$.  That is to say that the sum over radiating legs is written as $\sum\limits_{\ell \in X} $
where $X=\{q,\bar{q}\}$ or $X=\{g,g\}$, depending on the Higgs decay mode. When we turn to the results for specific processes in Section.~3, we will restore the process label.

\subsubsection{Sudakov Radiator}
Thrust is an additive global event-shape observable that describes the deviation of a final state from an exact two-particle configuration. 
In this case, the Sudakov radiator can be written up to NNLL as
\begin{equation}
  R_X(\lambda) = -\sum\limits_{\ell \in X} \frac{\lambda}{\alphas\beta_0}g_1^{(\ell)}(\lambda) + \left(g_2^{(\ell)}(\lambda) + h_2^{(\ell)}(\lambda)\right) + \frac{\alphas}{\pi}\left(g_3^{(\ell)}(\lambda) + \delta g_3^{(\ell)}(\lambda) + h_3^{(\ell)}(\lambda)\right)\,.
\end{equation}
For the thrust observable up to NLL, the radiator functions $g_1^{(\ell)}$, $g_2^{(\ell)}$, and $h_2^{(\ell)}$ are given by \cite{Catani:1991kz,Catani:1992ua}
\begin{align}
  g_1^{(\ell)}(\lambda) &= \frac{C_\ell}{2}\frac{2(1-\lambda)\log(1-\lambda) - (1-2\lambda)\log(1-2\lambda)}{\pi\beta_0\lambda}\,,\\
  \begin{split}
    g_2^{(\ell)}(\lambda) &= \frac{C_\ell}{2}\Bigg(\frac{K_1}{2\pi\beta_0^2}\frac{(\log(1-2\lambda) - 2\log(1-\lambda))}{\pi} + \frac{\beta_1}{\beta_0^3}\frac{\log(1-\lambda)^2}{\pi} \\
    &\qquad\qquad + \frac{\beta_1}{\beta_0^3}\frac{2\log(1-2\lambda)}{\pi} - \frac{\beta_1}{\beta_0^3}\frac{\log(1-2\lambda)(\log(1-2\lambda)+2)}{2\pi}\Bigg)\,,
  \end{split} \\
  h_2^{(\ell)}(\lambda) &= \frac{\gamma_\ell^{(0)}}{2\pi\beta_0}\log(1-\lambda)\,.
\end{align}
The soft radiator function $g_1^{(\ell)}$ contributes terms of order $\alphas^nL^{n+1}$, while the soft and hard radiator functions $g_2^{(\ell)}$ and $h_2^{(\ell)}$ contribute terms of the form $\alphas^nL^n$.
Their derivatives are given by
\begin{align}
  g_1^{\prime(\ell)}(\lambda) &= C_\ell\frac{\log(1-2\lambda)-2\log(1-\lambda)}{2\pi\beta_0\lambda^2}\,,\\
  g_1^{\prime\prime(\ell)}(\lambda) &= C_\ell\frac{-\log(1-2\lambda)+2\log(1-\lambda)}{\pi\beta_0\lambda^3} - C_\ell\frac{1}{\pi\beta_0(1-\lambda)(1-2\lambda)\lambda}\,,\\
  g_2^{\prime(\ell)}(\lambda) &= C_\ell\frac{2\pi\beta_1\left((1-\lambda)\log(1-2\lambda)-(1-2\lambda)\log(1-\lambda)\right)-(\beta_0K_1-2\pi\beta_1)\lambda}{2\pi^2\beta_0^3(1-\lambda)(1-2\lambda)}\,, \\
  h_2^{\prime(\ell)}(\lambda) &= -\frac{\gamma_\ell^{(0)}}{2\pi\beta_0(1-\lambda)}\,.
\end{align}
At NNLL, the soft radiator function $g_3^{(\ell)}$ and the hard-collinear radiator function $h_3^{(\ell)}$ are complemented by a ``mass correction'' $\delta g_3^{(\ell)}$, which arises from respecting the exact rapidity bound for emissions from a parton ensemble \cite{Banfi:2018mcq}.
The functions $g_3^{(\ell)}$, $h_3^{(\ell)}$, and $\delta g_3^{(\ell)}$ contribute corrections of order $\alphas^nL^{n-1}$.
For thrust, their specific form is given by \cite{Monni:2011gb,Banfi:2014sua,Banfi:2018mcq}
\begin{align}
  \begin{split}
    g_3^{(\ell)}(\lambda) &= \frac{C_\ell}{2}\Bigg(-\frac{K_2}{4\pi^2\beta_0^2}\frac{\lambda^2}{(1-2\lambda)(1-\lambda)} \\
    &\qquad\qquad + \frac{K_1\beta_1}{2\pi\beta_0^3}\frac{(6\lambda^2 + 2(1-\lambda)\log(1-2\lambda) - 4(1-2\lambda)\log(1-\lambda))}{2(1-2\lambda)(1-\lambda)}\\
    &\qquad\qquad - \frac{\beta_1^2}{\beta_0^4}\frac{\log(1-2\lambda)(4\lambda+\log(1-2\lambda))}{2(1-2\lambda)} \\
    &\qquad\qquad - \frac{\beta_1^2}{\beta_0^4}\frac{\lambda^2-(1-2\lambda)\log(1-\lambda)(2\lambda+\log(1-\lambda))}{(1-2\lambda)(1-\lambda)}\\
    &\qquad\qquad + \frac{\beta_2}{\beta_0^3}\frac{-2\lambda^2 + 2(1-2\lambda)(1-\lambda)(2\log(1-\lambda)-\log(1-2\lambda))}{2(1-2\lambda)(1-\lambda)}\Bigg)\,,
  \end{split}\\
  \delta g_3^{(\ell)}(\lambda) &= -C_\ell\zeta_2\frac{\lambda}{2(1-\lambda)}\,,\\
  h_3^{(\ell)}(\lambda) &= \gamma_\ell^{(0)}\frac{\beta_1(\log(1-\lambda)+\lambda)}{2\beta_0^2(1-\lambda)} - \gamma_\ell^{(1)}\frac{\lambda}{4\pi\beta_0(1-\lambda)}\,.
\end{align}
Throughout, $C_\ell$ denotes the quadratic Casimir associated to leg $\ell$, $C_\ell = \CA$ for gluons and $C_\ell = \CF$ for quarks.
The collinear anomalous dimension to order $n$ is denoted by $\gamma_\ell^{(n)}$.
Up to NNLL, only the first two orders are needed, which are given by 
\begin{align}
  \gamma_q^{(0)} &= -\frac{3}{2}\CF\,,\\
  \begin{split}
    \gamma_q^{(1)} &= -\frac{1}{2}\CF^2\left(\frac{3}{4}-\pi^2+12\zeta_3\right) - \frac{1}{2}\CF\CA\left(\frac{17}{12}+\frac{11}{9}\pi^2-6\zeta_3\right) \\
    &\qquad + \frac{1}{2}\CF\NF\left(\frac{1}{6}+\frac{2}{9}\pi^2\right)\,,
  \end{split}
\end{align}
for quarks, and by
\begin{align}
  \gamma_g^{(0)} &= -\frac{1}{6}\left(11\CA-2\NF\right)\,,\\
  \gamma_g^{(1)} &= -\CA^2\left(\frac{8}{3}+3\zeta_3\right) + \frac{2}{3}\CA\NF + \frac{1}{2}\CF\NF\,,
\end{align}
for gluons \cite{Ellis:1996mzs}.
The constants $K_1$ and $K_2$ are defined through the soft physical coupling $\alpha_\mathrm{s}^\text{phys}$, which generalises the CMW scheme \cite{Catani:1990rr} to the second order \cite{Banfi:2018mcq}.
Up to second order, $\alpha_\mathrm{s}^\text{phys}$ is related to the $\overline{\text{MS}}$ coupling by
\begin{equation}
  \alpha_\mathrm{s}^\text{phys} = \alpha_\mathrm{s}^{\overline{\text{MS}}}\left(1 + \frac{\alpha_\mathrm{s}^{\overline{\text{MS}}}}{2\pi}K_1 + \left(\frac{\alpha_\mathrm{s}^{\overline{\text{MS}}}}{2\pi}\right)^2 K_2\right)\,,
\end{equation}
with coefficients
\begin{align}
  K_1 &= \CA\left(\frac{67}{18}-\frac{\pi^2}{6}\right) - \frac{5}{9}\NF \,,\\
  \begin{split}
    K_2 &= \CA^2\left(\frac{245}{24} - \frac{67}{9}\zeta_2 + \frac{11}{6}\zeta_3 + \frac{11}{5}\zeta_2^2\right) + \CF\NF\left(-\frac{55}{24} + 2\zeta_3\right) \\
    &\qquad + \CA\NF\left(-\frac{209}{108} + \frac{10}{9}\zeta_2 - \frac{7}{3}\zeta_3\right) - \frac{1}{27}\NF^2 \\
    &\qquad + \frac{\pi\beta_0}{2}\left(\CA\left(\frac{808}{27} - 28\zeta_3\right) - \frac{224}{54}\NF\right)\,.
  \end{split}
\end{align}
While $K_1$ coincides with the two-loop cusp anomalous dimension, an analogous relationship for $K_2$ is only restored in the complete $\alphas^3L^2$ coefficient of the Sudakov radiator after an additional transverse-momentum integration \cite{Banfi:2018mcq}.

The following derivatives of the radiator functions will be required below
\begin{align}
  R_\text{NLL}'(\lambda) &= -\alphas\beta_0\frac{\rd}{\rd\lambda}\sum\limits_{\ell\in X}\frac{\lambda}{\alphas\beta_0}g_1^{(\ell)}(\lambda)\nonumber\\
  &= \frac{2C_\ell\left(\log(1-\lambda)-\log(1-2\lambda)\right)}{\pi\beta_0}\,,\\
  R_\text{NLL}''(\lambda) &= -(\alphas\beta_0)^2\frac{\rd^2}{\rd\lambda^2}\sum\limits_{\ell\in X}\frac{\lambda}{\alphas\beta_0}g_1^{(\ell)}(\lambda)\nonumber\\
  &= \frac{2C_\ell\alphas}{\pi(1-\lambda)(1-2\lambda)}\,,\\
  R_\text{NNLL}'(\lambda) &= -\alphas\beta_0\frac{\rd}{\rd\lambda}\sum\limits_{\ell\in X}g_2^{(\ell)}(\lambda)\nonumber\\
  \begin{split}
    &= \frac{C_\ell\alphas}{\pi^2\beta_0^2(1-\lambda)(1-2\lambda)}\Big(\left(K_1\beta_0 - 2\pi\beta_1\right)\lambda \\
    &\qquad\qquad + 2\pi\beta_1\left((1-2\lambda)\log(1-\lambda)\right)-(1-\lambda)\log(1-2\lambda)\Big)\,.
  \end{split}
\end{align}

\subsubsection{Multiple-Emission Function}
The multiple-emission function $\calF$ first enters at NLL, as only single emissions are considered at LL.
For additive observables, such as thrust, its NLL expression is given by \cite{Catani:1991kz,Catani:1992ua}
\begin{equation}
  \calF_\text{NLL}(\lambda) = \frac{\exp\left(-\gamma_\mathrm{E} R_\text{NLL}'(\lambda)\right)}{\Gamma\left(1+R_\text{NLL}'(\lambda)\right)}\,.
\end{equation}
Its derivative is given by
\begin{equation}
  \calF_\text{NLL}^\prime(\lambda) = -\frac{R_\text{NLL}''(\lambda)}{\alphas\beta_0}(\psi^{(0)}\left(1+R_\text{NLL}'(\lambda)\right)+\gamma_\mathrm{E})\calF_\text{NLL}(\lambda)\,.
\end{equation}
A general form for the NNLL multiple-emission function has been derived in \cite{Banfi:2014sua,Banfi:2018mcq}.
Specifically for thrust, it includes soft-collinear corrections $\delta\calF_\text{sc}$, hard-collinear corrections $\delta\calF_\text{hc}$, recoil corrections $\delta\calF_\text{rec}$, and correlated corrections $\delta\calF_\text{correl}$,
\begin{equation}
  \begin{split}
    \calF_\text{NNLL}(\lambda) &= \calF_\text{NLL}(\lambda)\left(1+\frac{\alphas}{2\pi}\frac{\lambda}{1-\lambda}\sum\limits_{\ell\in X}C_{\text{hc},\ell}^{(1)}\right) \\
    &\qquad + \frac{\alphas}{\pi}\Big(\delta\calF_\text{sc}(\lambda) + \delta\calF_\text{hc}(\lambda) + \delta\calF_\text{rec}(\lambda) + \delta\calF_\text{correl}(\lambda)\Big) \,.
  \end{split}
\end{equation}
We note that this formula differs from the one in \cite{Banfi:2018mcq} by non-logarithmic terms linear in $\alphas$.
This does not affect the formal logarithmic accuracy, but ensures that no spurious subleading terms are retained in the matching, see Sec.~\ref{subsec:matching}.
For thrust, the hard-collinear constants are
\begin{equation}
  C_{\text{hc},q}^{(1)} = \frac{9}{4}\CF\,,\qquad C_{\text{hc},g}^{(1)} = \frac{67}{36}\CA-\frac{7}{36}\NF \,,
\end{equation}
for quarks and gluons respectively \cite{Banfi:2018mcq,Arpino:2019ozn}.

The parametric form of the soft-collinear, hard-collinear, and correlated corrections are independent on the radiating flavour.
The soft-collinear term accounts for the exact rapidity boundary in a single soft-collinear emission and includes running-coupling corrections in the CMW scheme.
For thrust, it is given by \cite{Banfi:2014sua}
\begin{equation}
  \begin{split}
    \delta\calF_\text{sc}(\lambda) &= -\frac{\pi}{\alphas}\calF_\text{NLL}(\lambda)\Bigg[R_\text{NNLL}'(\lambda)\left(\psi^{(0)}(1+R_\text{NLL}'(\lambda))+\gamma_\mathrm{E}\right) \\
      &\qquad + \frac{R_\text{NLL}''(\lambda)}{2}\left(\left(\psi^{(0)}(1+R_\text{NLL}'(\lambda))+\gamma_\mathrm{E}\right)^2 - \psi^{(1)}(1+R_\text{NLL}'(\lambda)) + \frac{\pi^2}{6}\right)\Bigg]\,.
  \end{split}
\end{equation}
The correlated correction accounts for a single double-soft emission from an ensemble of independently emitted soft-collinear partons.
It is given for thrust by \cite{Banfi:2014sua}
\begin{equation}
  \delta\calF_\text{correl}(\lambda) = -\calF_\text{NLL}(\lambda)\frac{\lambda R_\text{NLL}''(\lambda)}{2\alphas\beta_0}\left(\pi\beta_0\zeta_2 - \frac{11}{6}\zeta_2\CA + \frac{1}{3}\zeta_2\NF\right)\,.
\end{equation}
The hard-collinear correction arises from the effect of a hard-collinear emission on the squared matrix element.
For thrust, it reads \cite{Banfi:2014sua}
\begin{equation}
  \delta\calF_\text{hc}(\lambda) = -\calF_\text{NLL}(\lambda)\sum\limits_{\ell=1}^{2} \gamma_\ell^{(0)}\left(\psi^{(0)}\left(1+R_\text{NLL}'(\lambda)\right) + \gamma_\text{E}\right)\frac{1}{2(1-\lambda)}\,.
\end{equation}

The recoil correction is derived using the method outlined in \cite{Banfi:2014sua} yielding
\begin{equation}
  \delta\calF_{\text{rec},q\bar{q}}(\lambda) = \CF\left(\frac{5}{4}-\frac{\pi^2}{3}\right)\frac{\lambda}{1-\lambda}\calF_\text{NLL}(\lambda)\,,
\end{equation}
for $q\bar{q}$ configurations, and
\begin{equation}
  \delta\calF_{\text{rec},gg}(\lambda) = \left(\left(\frac{67}{36}-\frac{\pi^2}{3}\right)\CA-\frac{13}{36}\NF\right)\frac{\lambda}{1-\lambda}\calF_\text{NLL}(\lambda)\,,
\end{equation}
for $gg$ configurations.
These formulae again differ from the ones quoted in \cite{Banfi:2018mcq,Arpino:2019ozn} by non-logarithmic terms linear in $\alphas$ to simplify the matching procedure below.

\begin{figure}[t]
  \centering
  \includegraphics[width=0.48\textwidth]{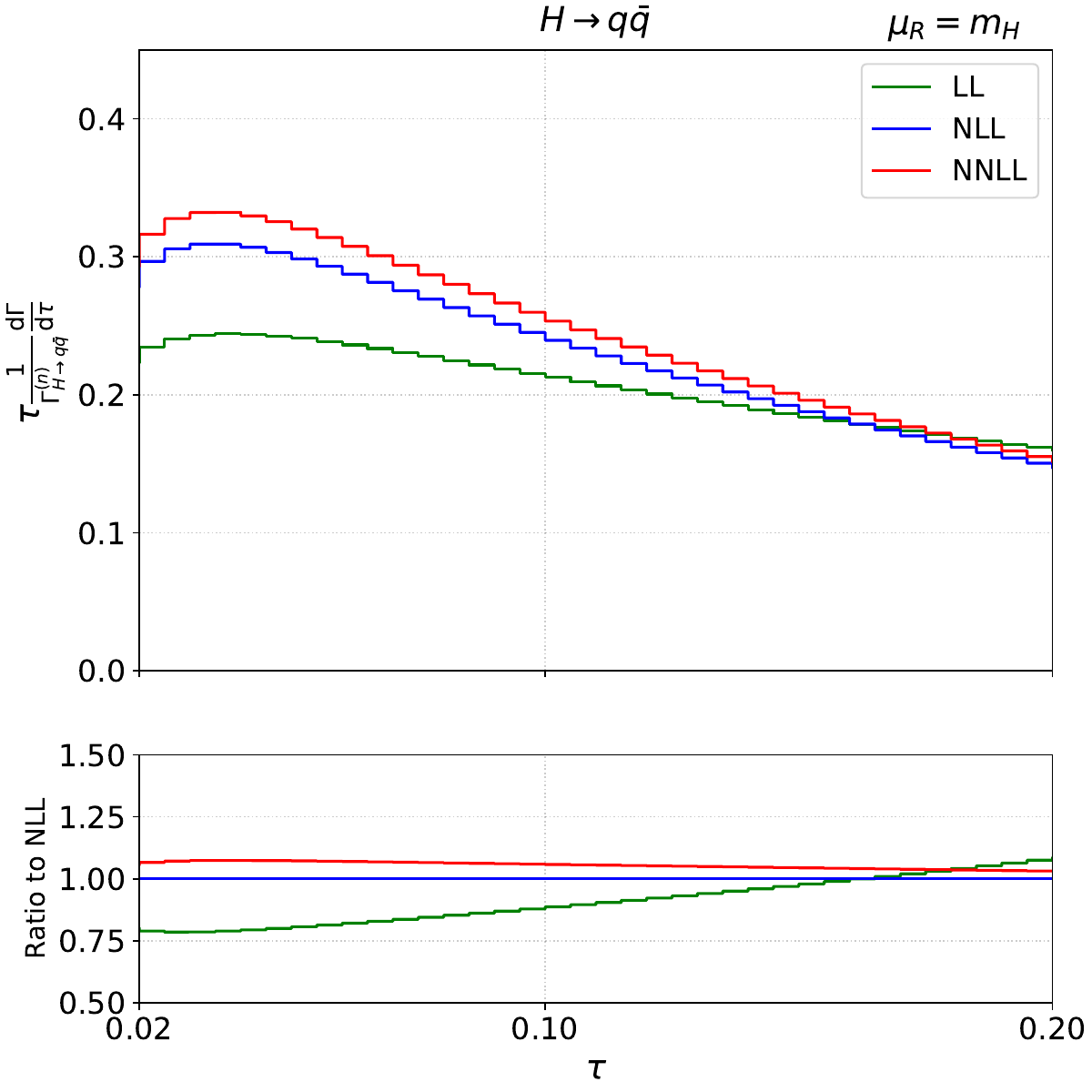}
  \includegraphics[width=0.48\textwidth]{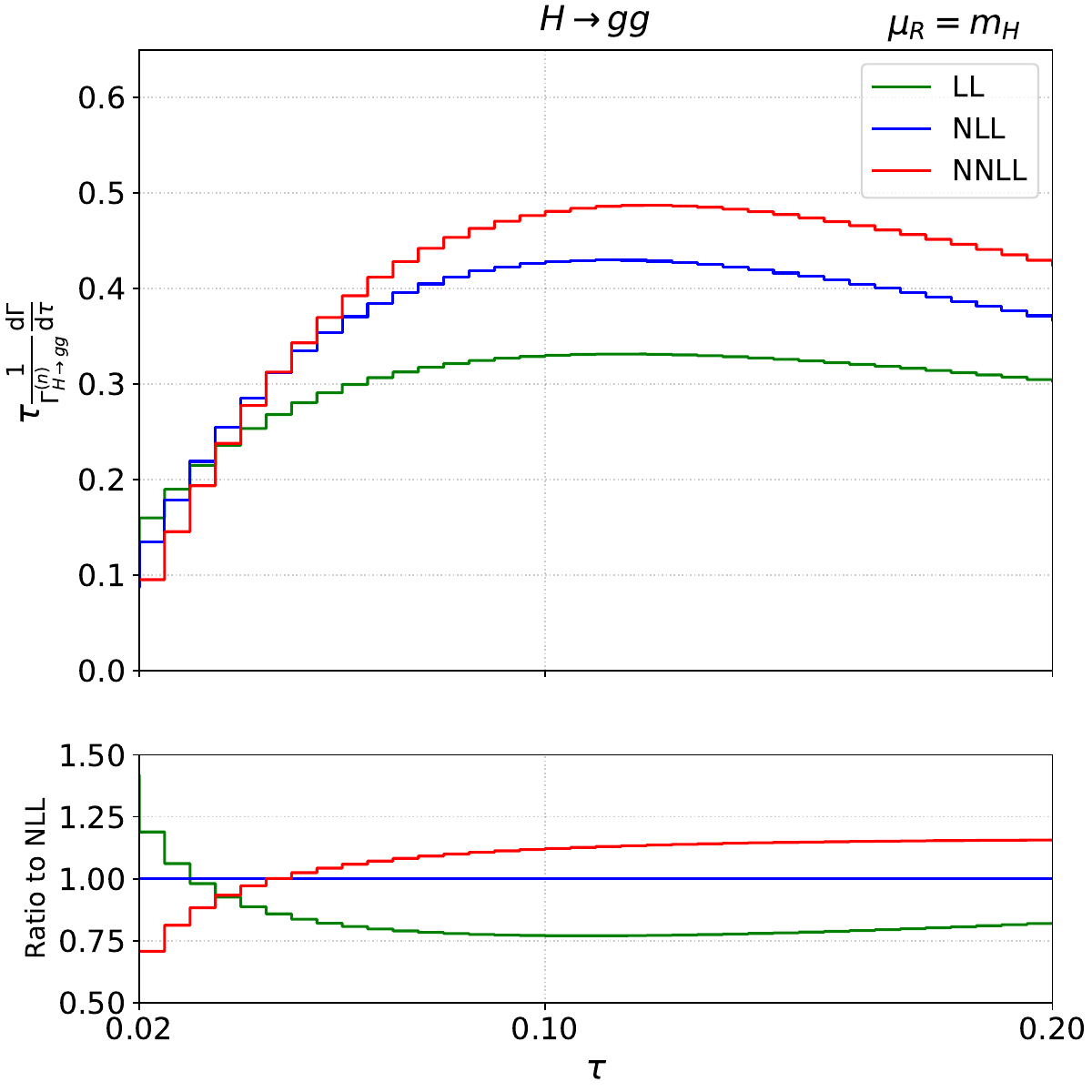}
  \caption{Comparison of resummed results for thrust at LL (green), NLL (blue), and NNLL (red) in $H\to q\bar{q}$ decays (left) and $H\to gg$ decays (right). }
  \label{fig:resummed}
\end{figure}

Fig.~\ref{fig:resummed} shows a comparison of the LL, NLL, and NNLL results in the $H\to q\bar{q}$ and \mbox{$H\to gg$} channels in the peak region, with $\muR=m_H=125.09~\mathrm{GeV}$ and $\alpha_s(\muR)=0.11263$.
As expected from naive Casimir-scaling arguments, we observe a significant shift of the peak of the distributions towards higher values of $\tau$ in the $H\to gg$ decay mode.
Nevertheless, the behaviour is qualitatively the same between $H\to q\bar{q}$ and $H\to gg$.
The NNLL corrections are sizeable in both decay channels, albeit larger in $H\to gg$. At the peak, they amount to $+8$\% and  $+10$\% with respect to the NLL for the $H\to q\bar{q}$ and $H\to gg$ channels respectively. As $\tau$ increases beyond the peak region, the relative size of the NNLL corrections decreases for the $H\to q\bar{q}$ channel, while for the $H\to gg$ channel it increases.

\subsection{Matching NNLO to NNLL}
\label{subsec:matching}
The fixed-order expansion of the cumulative cross section \eqref{eq:cumulant} to third order in the strong coupling reads
\begin{equation}
  \Sigma(\tau) = 1 + \frac{\alphas}{2\pi}\mathcal{A}(\tau) + \left(\frac{\alphas}{2\pi}\right)^2\mathcal{B}(\tau) + \left(\frac{\alphas}{2\pi}\right)^3\mathcal{C}(\tau) + \order{\alphas^4}\,,
\end{equation}
where the coefficients $\mathcal{A}$, $\mathcal{B}$, and $\mathcal{C}$ can be obtained from the LO, NLO, and NNLO coefficients $A$, $B$, and $C$ in \eqref{eq:rateDiff}.
As the matching procedure is generic, we suppress any explicit labels of $X$ and $H\to X$.
Its logarithmic dependence is given by
\begin{align}
  \mathcal{A}(\tau) &= c_1 + G_{11}L + G_{12}L^2 + D_1(\tau) \,, \label{eq:logDepA}\\
  \begin{split}
    \mathcal{B}(\tau) &= c_2 + \left(G_{21}+c_1G_{11}\right)L + \left(G_{22}+\frac{1}{2}G_{11}^2+c_1G_{12}\right)L^2 \\
    &\qquad \left(G_{23}+G_{12}G_{11}\right)L^3 + \frac{1}{2}G_{12}^2L^4 + D_2(\tau) \,,
  \end{split} \label{eq:logDepB} \\
  \begin{split}
    \mathcal{C}(\tau) &= c_3 + \left(G_{31}+c_1G_{21}+c_2G_{11}\right)L \\
    &\qquad + \left(G_{32}+c_1G_{22}+\frac{1}{2}c_1^2G_{11}+c_2G_{12}+G_{11}G_{21}\right)L^2 \\
    &\qquad + \left(G_{33}+G_{11}G_{22}+G_{12}G_{21}+c_1G_{11}G_{12}+\frac{1}{6}G_{11}^3+c_1G_{23}\right)L^3 \\
    &\qquad + \left(G_{34}+G_{12}G_{22}+\frac{1}{2}c_1G_{12}^2+G_{11}G_{23}+\frac{1}{2}G_{11}^2G_{11}\right)L^4 \\
    &\qquad + \left(G_{12}G_{23}+\frac{1}{2}G_{12}^2G_{11}\right)L^5 + \frac{1}{6}G_{12}L^6 + D_3(\tau) \,.
  \end{split} \label{eq:logDepC}
\end{align}
The functions $D_k(\tau)$, $k=1,2,3$, contain power corrections 
in $\tau$ from the fixed-order prediction at perturbative order $k$.
In particular, $D_k(\tau) \to 0$ for $\tau \to 0$.
Following the definition of the logarithmic counting in the exponent, the coefficients $G_{mn}$ arise from the expansion of the logarithm of the cumulant as
\begin{align}
  \Sigma_\text{LL}(\tau) &= \exp\left\{G_{12}\left(\frac{\alphas}{2\pi}\right)L^2 + G_{23}\left(\frac{\alphas}{2\pi}\right)^2L^3 + G_{34}\left(\frac{\alphas}{2\pi}\right)^3L^4 + \ldots\right\} \,, \\
  \Sigma_\text{NLL}(\tau) &= \exp\left\{G_{11}\left(\frac{\alphas}{2\pi}\right)L + G_{22}\left(\frac{\alphas}{2\pi}\right)^2L^2 + G_{33}\left(\frac{\alphas}{2\pi}\right)^3L^3 + \ldots\right\} \,, \\
  \Sigma_\text{NNLL}(\tau) &= \exp\left\{G_{21}\left(\frac{\alphas}{2\pi}\right)^2L + G_{32}\left(\frac{\alphas}{2\pi}\right)^3L^2 + \ldots\right\} \,.
\end{align}
It is evident that the NNLL resummation does not capture the single-log term $G_{31}$ in the NNLO coefficient $\mathcal{C}$.
This would require including terms of order $\alphas^nL^{n-2}$, which only appear at \N{3}LL. The expressions for the $G_{ij}$ coefficients are given in App.~\ref{app:C}.

In the following, we will consider the logR-matching scheme \cite{Catani:1992ua,Jones:2003yv}.
At NNLO+NNLL, it reads in our notation
\begin{equation}
  \begin{split}
    \log\left(\Sigma(\tau)\right) &= -R_\text{NNLL}(\tau) + \log\left(\calF_\text{NNLL}(\tau)\right) + \frac{\alphas}{2\pi}\left(\mathcal{A}(\tau) - G_{11}L - G_{12}L^2\right) \\
    &\quad + \left(\frac{\alphas}{2\pi}\right)^2\left(\mathcal{B}(\tau) - \frac{1}{2}\mathcal{A}(\tau)^2 - G_{21}L - G_{22}L^2 - G_{23}L^3\right) \\
    &\quad + \left(\frac{\alphas}{2\pi}\right)^3\left(\mathcal{C}(\tau) - \mathcal{A}(\tau)\mathcal{B}(\tau) + \frac{1}{3}\mathcal{A}(\tau)^3 - G_{32}L^2 - G_{33}L^3 - G_{34}L^4\right)\,. \\
  \end{split}
\end{equation}
Notably, the $G_{31}$ coefficient does not explicitly enter the logR scheme.
The physical behaviour of $\Sigma(\tau_\mathrm{max})\to 1$ as $\tau \to \tau_\mathrm{max}$ is recovered by modifying the large logarithm $L$ as \cite{Jones:2003yv}
\begin{equation}
  L \to L' = \log\left(\left(\frac{1}{\tau}\right) - \left(\frac{1}{\tau_\mathrm{max}}\right) + 1\right) \,.
  \label{eq:modifiedLog}
\end{equation}
At LO and NLO, the kinematical endpoints can easily be derived from geometrical arguments as
\begin{equation}
  \tau_\mathrm{max}^{(\LO)}=\frac{1}{3} \,,\qquad \tau_\mathrm{max}^{(\NLO)}=1-\frac{1}{\sqrt{3}}\,.
\end{equation}
At NNLO, we determine the kinematical endpoint numerically as $\tau_\mathrm{max}^{(\NNLO)} = 0.4349$.

We assess theoretical uncertainties through variations of the renormalisation scale $\muR$ by a factor $x_\mu$.
The renormalisation-scale dependence is extracted by considering the running of the strong coupling,
\begin{equation}
  \begin{split}
    \alphas(x_\mu\muR) &= \frac{\alphas(\muR)}{1+\alphas(\muR)\beta_0\log(x_\mu^2)} \\
    &\qquad \times \left(1 - \frac{\alphas(\muR)}{1+\alphas(\muR)\beta_0\log(x_\mu^2)}\frac{\beta_1}{\beta_0}\log(1+\alphas(\muR)\beta_0\log(x_\mu^2))\right) \,,
  \end{split}
\end{equation}
and keeping only $\order{\alphas}$ terms relative to the coupling order \cite{Monni:2011gb}.
Moreover, we also vary the choice of logarithms to be resummed by introducing a parametrisation $\tau \to x_L \tau$,
\begin{equation}
  -\log(x_L \tau) = L - \log(x_L)\,.
\end{equation}
We further require $\Sigma(\lambda,x_\mu,x_L) \to 1$ as $\lambda\to 0$, so that
\begin{equation}
  g_2^{(\ell)}(0) = 0\,,\qquad g_3^{(\ell)}(0) = 0\,, \qquad h_3^{(\ell)}(0) = 0 \,.
\end{equation}
The combined $(x_\mu,x_L)$ scale variations then yield the following radiator functions,
\begin{align}
  g_2^{(\ell)}(\lambda,x_L,x_\mu) &= g_2^{(\ell)}(\lambda) + \left(g_1^{(\ell)}(\lambda) + \lambda g_1^{\prime(\ell)}(\lambda)\right)\log(x_L) + \lambda^2 g_1^{\prime(\ell)}(\lambda)\log(x_\mu^2)\,, \label{eq:scalevar1} \\ 
  \begin{split}
    g_3^{(\ell)}(\lambda,x_L,x_\mu) &= g_3^{(\ell)}(\lambda) + \frac{C_\ell}{2}\log(x_L)^2 \\
    &\quad + \pi\beta_0\left(g_2^{\prime(\ell)}(\lambda)\left(\log(x_L) + \lambda\log(x_\mu^2)\right) + \frac{\lambda^2\beta_1}{\beta_0^2}g_1^{\prime(\ell)}(\lambda)\log(x_\mu^2)\right) \\
    &\quad + \pi\beta_0\left(g_1^{\prime(\ell)}(\lambda) + \frac{1}{2}\lambda g_1^{\prime\prime(\ell)}(\lambda)\right)\left(\lambda^2\log(x_\mu^2)^2 + \log(x_L)^2\right) \\
    &\quad + \pi\beta_0\lambda\left(2g_1^{\prime(\ell)}(\lambda) + \lambda g_1^{\prime\prime(\ell)}(\lambda)\right)\log(x_L)\log(x_\mu^2)  
  \end{split} 
  \label{eq:scalevar2} \\
  \begin{split}
    h_3^{(\ell)}(\lambda,x_L,x_\mu) &= h_3^{(\ell)}(\lambda) + \frac{\gamma_\ell^{(0)}}{2}\log(x_L)\\
    &\quad + \pi\beta_0 h_2^{\prime(\ell)}(\lambda)\log(x_L) + \pi\beta_0\lambda h_2^{\prime(\ell)}(\lambda)\log(x_\mu^2)\,,
  \end{split} 
    \label{eq:scalevar3}
\end{align}
while the scale variation of the the NNLL multiple-emission function reads
\begin{equation}
  \begin{split}
    \calF_\text{NNLL}(\lambda,x_L,x_\mu) &= \Bigg(\calF_\text{NLL}(\lambda) + \alphas\beta_0\log(x_L)\calF_\text{NLL}^\prime(\lambda) + \alphas\beta_0\log(x_\mu^2)\calF_\text{NLL}^\prime(\lambda)\Bigg)\\
    &\qquad \times\left(1+\frac{\alphas}{2\pi}\frac{\lambda}{1-\lambda}\sum\limits_{\ell\in X}C_{\text{hc},\ell}^{(1)}\right) \\
    &\qquad + \frac{\alphas}{\pi}\Big(\delta\calF_\text{sc}(\lambda) + \delta\calF_\text{hc}(\lambda) + \delta\calF_\text{rec}(\lambda) + \delta\calF_\text{correl}(\lambda)\Big) \,.
  \end{split}
\end{equation}

\section{Results}
\label{sec:results}

Based on the framework described in the previous section, we can derive predictions
for the thrust distribution in the Higgs decays $H\to q\bar{q}$ and $H\to gg$. We first discuss the analytical extraction of 
the $c_2$ and $G_{31}$ coefficients in Sec.~\ref{sec:results_extraction}. Secondly, we present a numerical validation of the  fixed-order and resummed calculations in Sec.~\ref{sec:results_validation}, and the complete results for matched predictions of the thrust distribution at NNLO+NNLL in Sec.~\ref{sec:results_matched}.

\subsection{Analytical Extraction of $c_2$ and $G_{31}$}
\label{sec:results_extraction}
The analytical knowledge of the thrust distribution at NNLL allows us to derive an analytical value for the $c_2$ parameter in \eqref{eq:logDepB} from the leading singular terms calculated in \cite{Gao:2019mlt}.
Specifically, we derive the $\alphas^3L^2$ coefficient from the third-order expansion of the NNLL cumulant cross section and cast it in the form
\begin{equation}
  G_{32} + c_1G_{22} + \frac{1}{2}c_1^2G_{11} + c_2G_{12} + G_{11}G_{21} \,.
\end{equation}
An equivalent form can be derived by integrating the leading-singular terms of the differential thrust distribution in the $H\to q\bar{q}$ and $H\to gg$ decay channels given in \cite{Gao:2019mlt}, albeit without identification of LL, NLL, and NNLL terms.
Except for $c_2$, the $\alphas^3L^2$ term contains at most NNLL terms, so that $c_2$ can be derived exactly from the correspondence with the integrated result of \cite{Gao:2019mlt}.
For $H\to q\bar{q}$ we find
\begin{equation}
  \begin{split}
    c_2^{(H\to q\bar{q})} &= \left(\frac{8474}{81}-\frac{9398}{243}\pi^2+\frac{23026}{81}\zeta_3+\frac{163}{270}\pi^4\right) \\
    &\quad + \left(-\frac{745}{18}+\frac{73}{9}\pi^2-\frac{3548}{81}\zeta_3-\frac{19}{405}\pi^4\right)\NF \\
    &\quad + \left(\frac{191}{162}-\frac{61}{243}\pi^2+\frac{32}{27}\zeta_3\right)\NF^2\,,
  \end{split}
\end{equation}
while for $H\to gg$ this yields
\begin{equation}
  \begin{split}
    c_2^{(H\to gg)} &= \left(-\frac{6775}{24}-\frac{487}{24}\pi^2+\frac{627}{2}\zeta_3+\frac{61}{40}\pi^4\right) \\
    &\quad + \left(\frac{149}{4}+\frac{3}{2}\pi^2-\frac{67}{3}\zeta_3\right)\NF + \left(-\frac{25}{27}+\frac{1}{54}\pi^2\right)\NF^2\,.
  \end{split}
\end{equation}

Once the $c_2$ coefficient is determined, it is further possible to extract the \N3LL coefficient $G_{31}$ from the $\alphas^3L$ term,
\begin{equation}
  G_{31} + c_1G_{21} + c_2G_{11} \,.
\end{equation}
Using the results from \cite{Gao:2019mlt} we find in the $q\bar{q}$ mode,
\begin{equation}
  \begin{split}
    G_{31,q\bar{q}} &= \left(\frac{37594}{27} - \frac{6223}{27}\pi^2 + \frac{25168}{27}\zeta_3 + \frac{391}{405}\pi^4 - \frac{1952}{81}\pi^2\zeta_3 + \frac{5336}{9}\zeta_5\right) \\
    &\quad + \left(-\frac{4934}{27} + \frac{2500}{81}\pi^2 - \frac{12176}{81}\zeta_3 - \frac{76}{405}\pi^4\right)\NF \\
    &\quad + \left(\frac{382}{81} - \frac{244}{243}\pi^2 + \frac{128}{27}\zeta_3\right)\NF^2 \,,
  \end{split}
  \label{eq:G31qqb}
\end{equation}
which coincides with Eq.~(4.33) of~\cite{Monni:2011gb} for $\CA=3$ and $\CF=4/3$, while in the $gg$ decay mode we find
\begin{equation}
  \begin{split}
    G_{31,gg} &= \left(\frac{312013}{72} - \frac{11399}{24}\pi^2 + 4887\zeta_3 - \frac{627}{20}\pi^4 - 342\pi^2\zeta_3 + 4644\zeta_5\right) \\
    &\quad + \left(-\frac{5921}{8} + \frac{2863}{36}\pi^2 + \frac{19}{10}\pi^4 - \frac{2428}{3}\zeta_3\right)\NF \\
    &\quad + \left(\frac{2303}{72} - \frac{229}{54}\pi^2 + \frac{268}{9}\zeta_3\right)\NF^2 + \left(-\frac{50}{243} + \frac{5}{81}\pi^2\right)\NF^3 \,.
  \end{split}
  \label{eq:G31gg}
\end{equation}

\subsection{Numerical Validation in the Infrared Limit}
\label{sec:results_validation}
We start by validating our implementations in the two-particle limit by numerically comparing the LO, NLO, and NNLO fixed-order predictions against the $\order{\alphas}$, $\order{\alphas^2}$, and $\order{\alphas^3}$ expansion of the resummed prediction.
This provides a valuable cross check of both implementations.

\begin{figure}[t]
  \centering
  \includegraphics[width=0.48\textwidth]{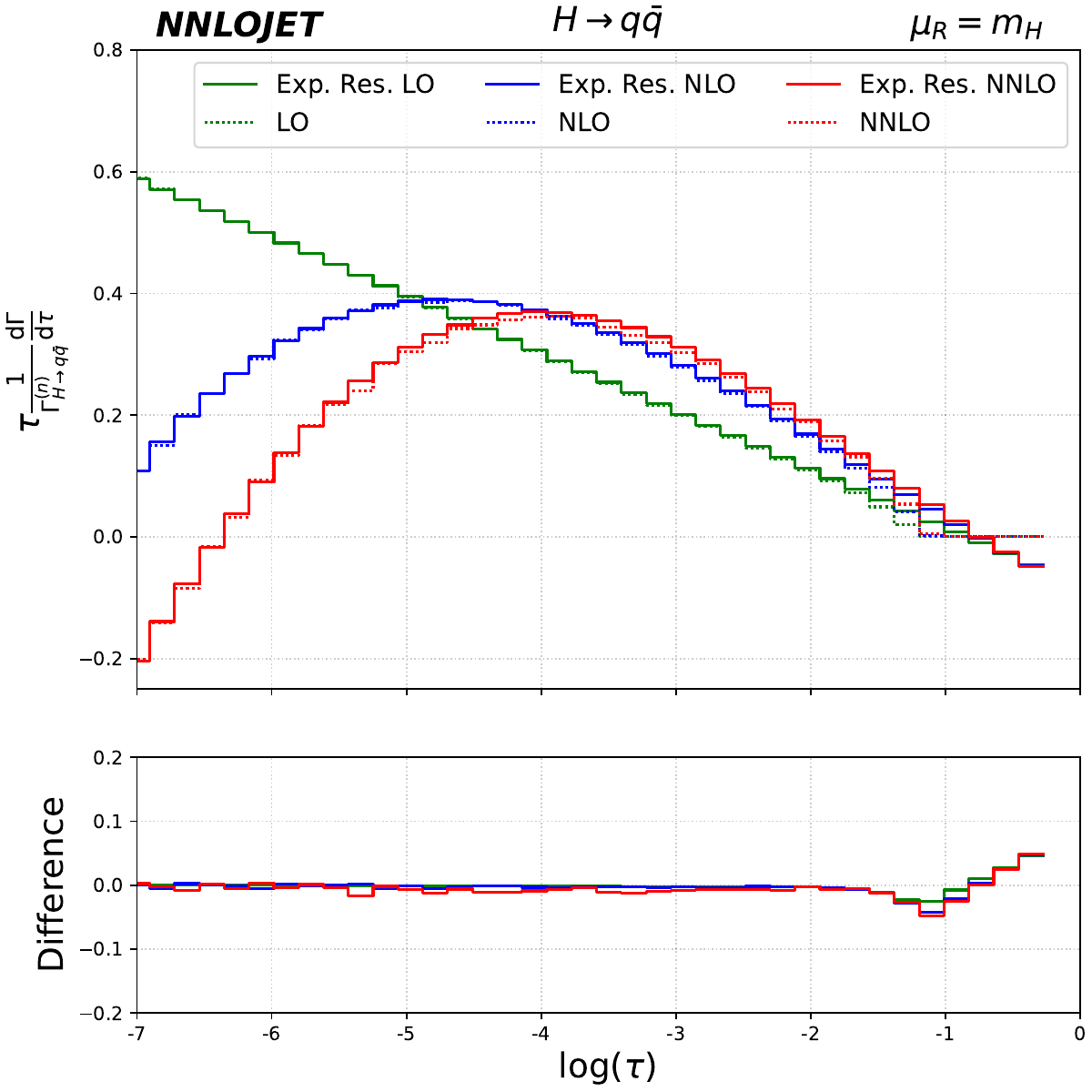}
  \includegraphics[width=0.48\textwidth]{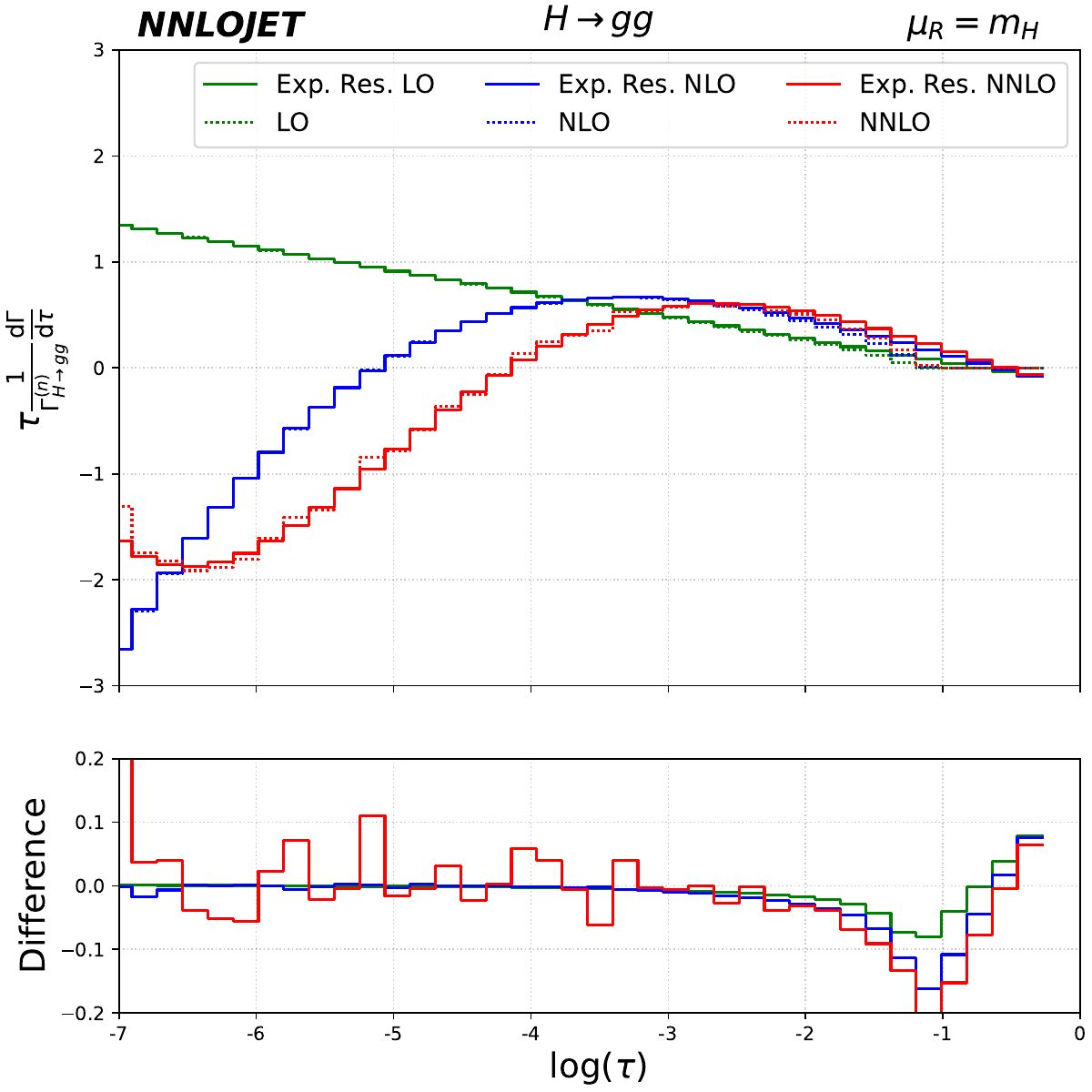}
  \caption{Comparison between the expansion of the resummation formula (solid lines) and the fixed-order results (dashed lines) up to $\mathcal{O}(\alpha_s)$ (LO, green), up to $\mathcal{O}(\alpha_s^2)$ (NLO, blue) and up to  $\mathcal{O}(\alpha_s^3)$ (NNLO, red). The difference between  the fixed-order and the expansion of the resummation formula is shown in the lower frames.}
  \label{fig:check}
\end{figure}

Fig.~\ref{fig:check} contains the comparison of the fixed-order and expanded resummed predictions in the limit $\tau \to 0$.
The larger main panels contain the fixed-order predictions of the differential decay rate at LO, NLO, and NNLO compared to the first-order (LO), second-order (NLO), and third-order (NNLO) expansion of the NNLL prediction.
The smaller panels below the main ones show the difference between the LO, NLO, and NNLO prediction and the respective fixed-order expansion of the resummed result.
In both the fixed-order and resummed predictions, the renormalisation scale is set to $\muR=m_H=125.09~\mathrm{GeV}$, so that $\alpha_s(\mu_R)=0.11263$.

We observe excellent convergence below $\log(\tau) \approx -2$ in both the $H\to q\bar{q}$ and $H\to gg$ decay modes at LO and NLO.
This is because the full logarithmic structure of the fixed-order prediction is captured by the NNLL prediction.
However, at NNLO, the NNLL result does not capture the full $\alphas^3L$ coefficient, due to the missing \N3LL factor $G_{31}$.
Upon amending the NNLL expansion by this term (included in the results in Fig.~\ref{fig:check}), we observe very good agreement with the NNLO result in the region $\log(\tau) < -2$, as shown in the plots.
The values of $G_{31}$ in the $H\to b\bar{b}$ and $H\to gg$ channels are given in Sec.~\ref{sec:results_extraction}.

\subsection{Matched Predictions at NNLO+NNLL}
\label{sec:results_matched}
In this section, we provide matched predictions of the thrust observable in the \mbox{$H\to b\bar{b}$}, \mbox{$H\to c\bar{c}$}, and $H\to gg$ decay modes at NNLO+NNLL.
We consider all electroweak parameters as constant and set them in the $G_\upmu$ scheme, with input values
\begin{equation}
  \GF =  1.1664\times 10^{-5}~\GeV^{-2}\,, \quad m_Z = 91.200~\GeV \,, 
\end{equation}
where the Fermi constant $\GF$ determines the (effective) $Hgg$ and $Hq\bar{q}$ couplings $\lambda_0$ and $y_q$.
The vacuum expectation value is given by $v=({\sqrt{2}\GF})^{-\tfrac{1}{2}}=246.22$ GeV and the $Z$-boson mass $m_Z$ serves as the reference scale for the strong coupling.
The central scale is chosen as $\muR = m_H = 125.09~\GeV$ and perturbative uncertainties are assessed by combined scale variations of the renormalisation and resummation scales, $\muR\to x_\mu \muR$ and $\tau \to x_L \tau$ with $x_\mu, x_L \in \left[\frac{1}{2},2\right]$
The nominal value of the strong coupling at the $Z$-boson mass is set to $\alphas(m_Z) = 0.11800$ and the strong coupling is evaluated at one, two, or three loops at LO, NLO, and NNLO, respectively.

Below, we present differential distributions
\begin{equation}
  \tau\frac{1}{\Gamma_{H\to X}^{(k)}(m_H,\mu_R)}\frac{\rd \Gamma_{H\to X}(m_H,\mu_R)}{\rd \tau}\,,
\end{equation}
including the Yukawa-induced decay to quarks and the decay to gluons.
For the latter decay mode, the effective coupling $\lambda_0^2(\mu_R)$ in~\eqref{eq:couplings} is rescaled to include finite top, bottom, and charm mass effects~\cite{Spira:1997dg}, as well as electroweak corrections~\cite{Actis:2008ug}.
We also present differential distributions of the sum of decay channels, where we consider the decay to bottom and charm quarks, which has a significant impact on phenomenological predictions.
The \mbox{$H\to b\bar{b}$} and $H\to c\bar{c}$ decays are formally identical, with the latter obtained from the former by applying a rescaling factor $y_c^2(\muR)/y_b^2(\muR)$.
The total hadronic decay width at a given perturbative order $k$ is given by
\begin{equation}
  \Gamma^{(k)}=\Gamma^{(k)}_{H\to b\bar{b}}+\Gamma^{(k)}_{H\to c\bar{c}}+\Gamma^{(k)}_{H\to gg}\,,
\end{equation}
so that the sum over decay modes at that order is defined as
\begin{equation}
  \tau\frac{1}{\Gamma^{(k)}(m_H,\mu_R)}\sum_{X}\frac{\rd \Gamma_{H\to X}(m_H,\mu_R)}{\rd \tau}, \quad\text{with}\,X=b\bar{b},c\bar{c},gg \,.
\end{equation}
The bottom- and charm-quark Yukawa couplings are considered to be running in the $\overline{\text{MS}}$ scheme with \mbox{$y_b(m_H)=m_b(m_H)/v=0.011309$}, \mbox{$y_c(m_H)=m_c(m_H)/v=0.0024629$}.
The $\overline{\text{MS}}$ top-quark mass is chosen as \mbox{$m_t(m_H) = 166.48~\GeV$}.
Quark masses and Yukawa couplings are evolved following the results of~\cite{Vermaseren:1997fq}.

\begin{figure}[t]
  \centering
  \includegraphics[width=0.48\textwidth]{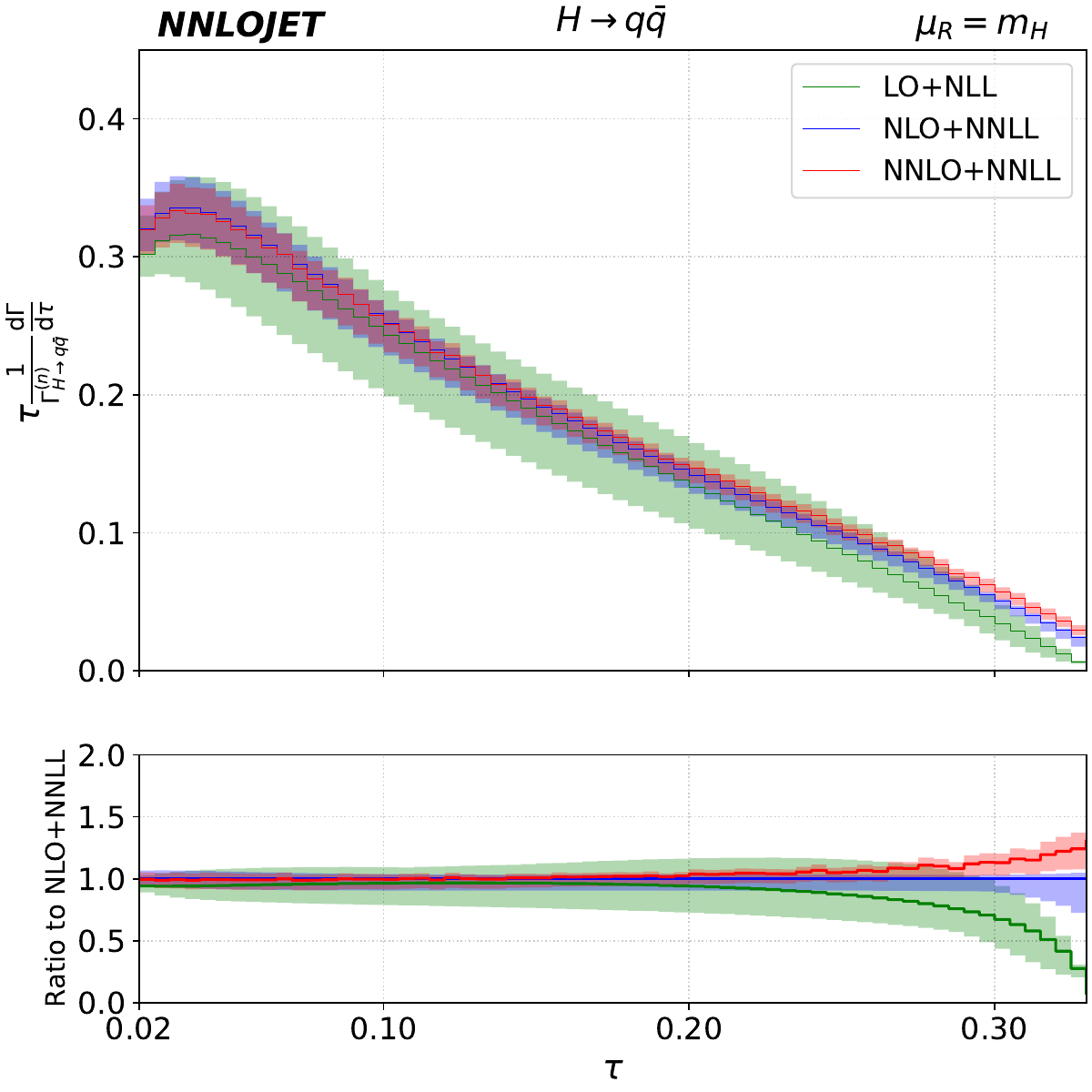}
  \includegraphics[width=0.48\textwidth]{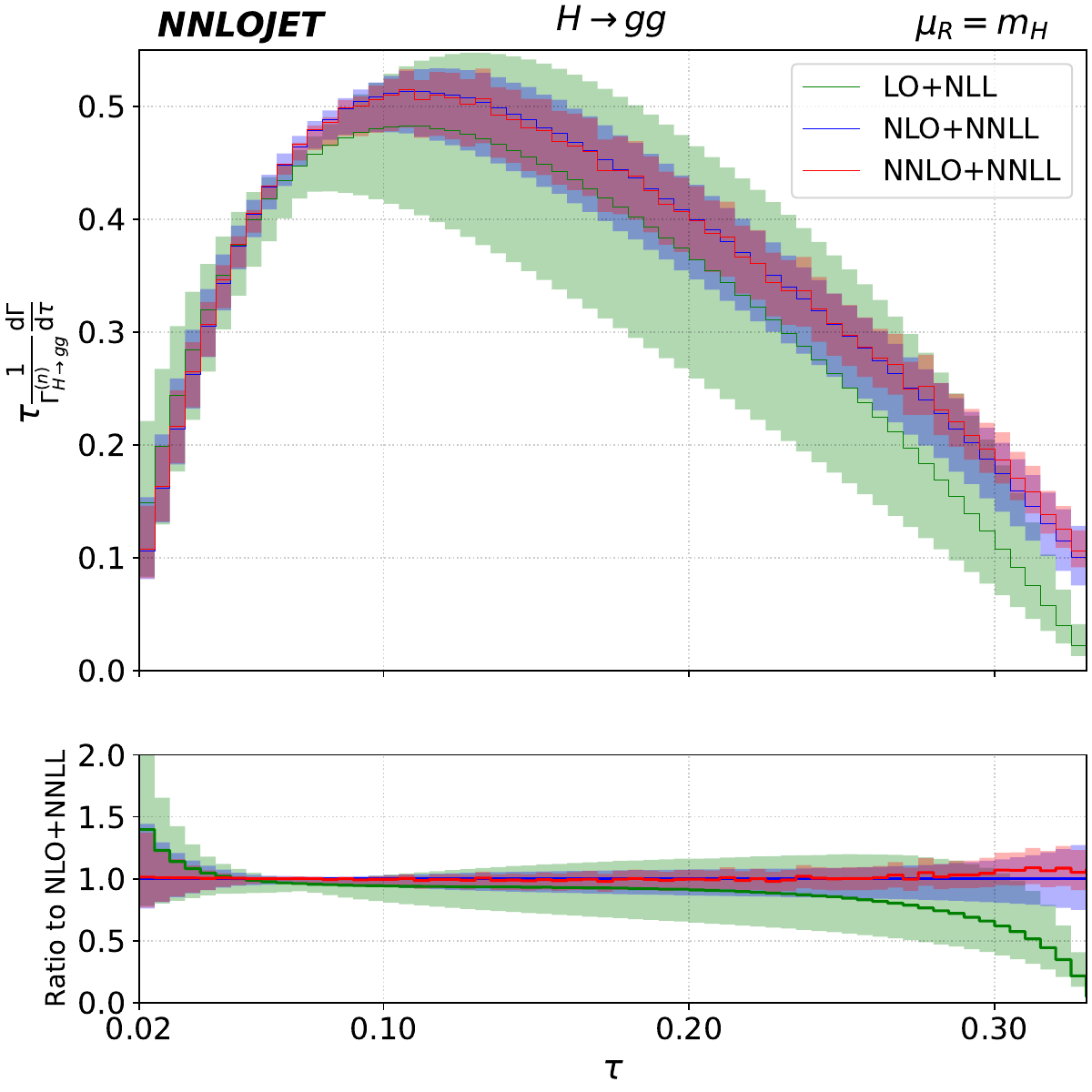}
  \includegraphics[width=0.48\textwidth]{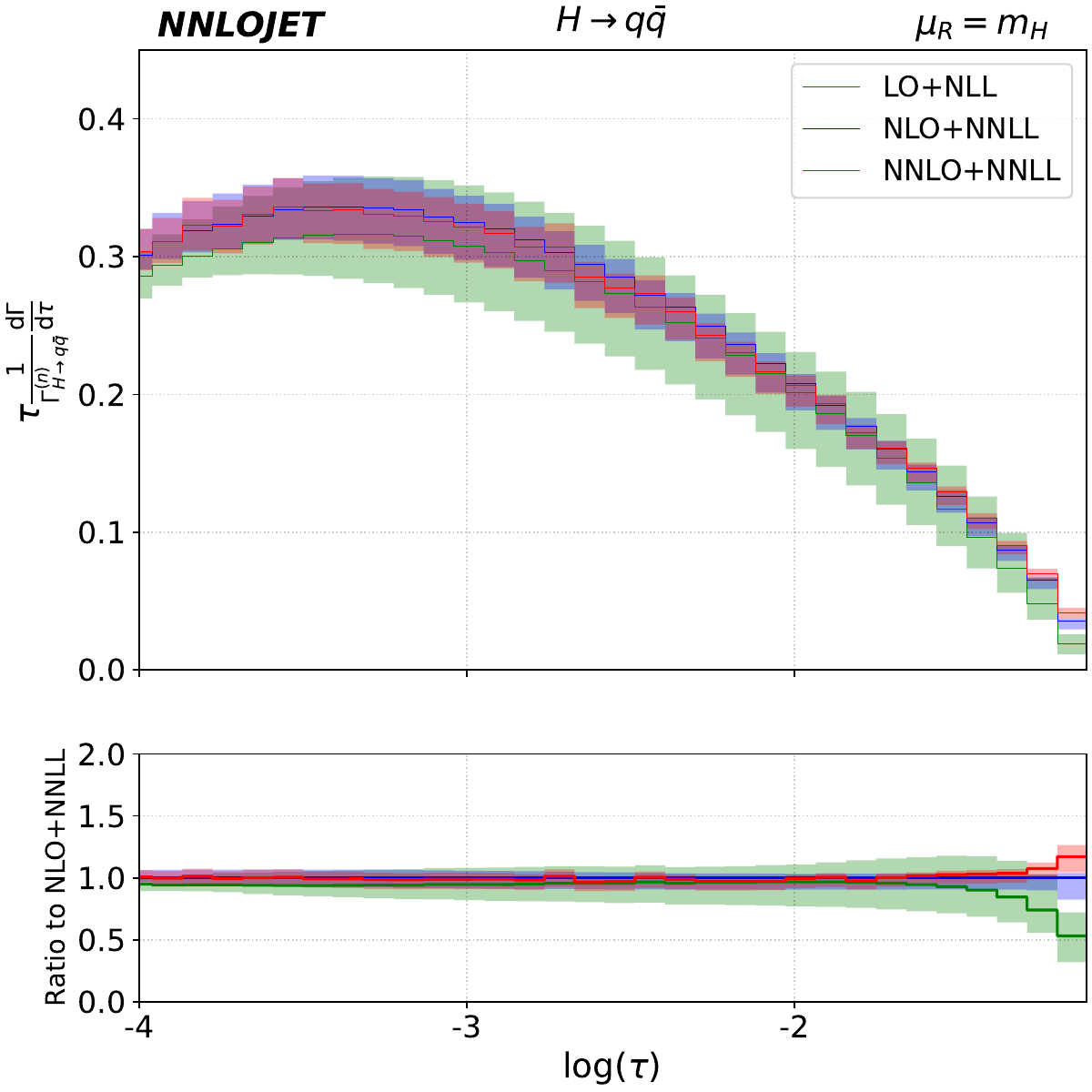}
  \includegraphics[width=0.48\textwidth]{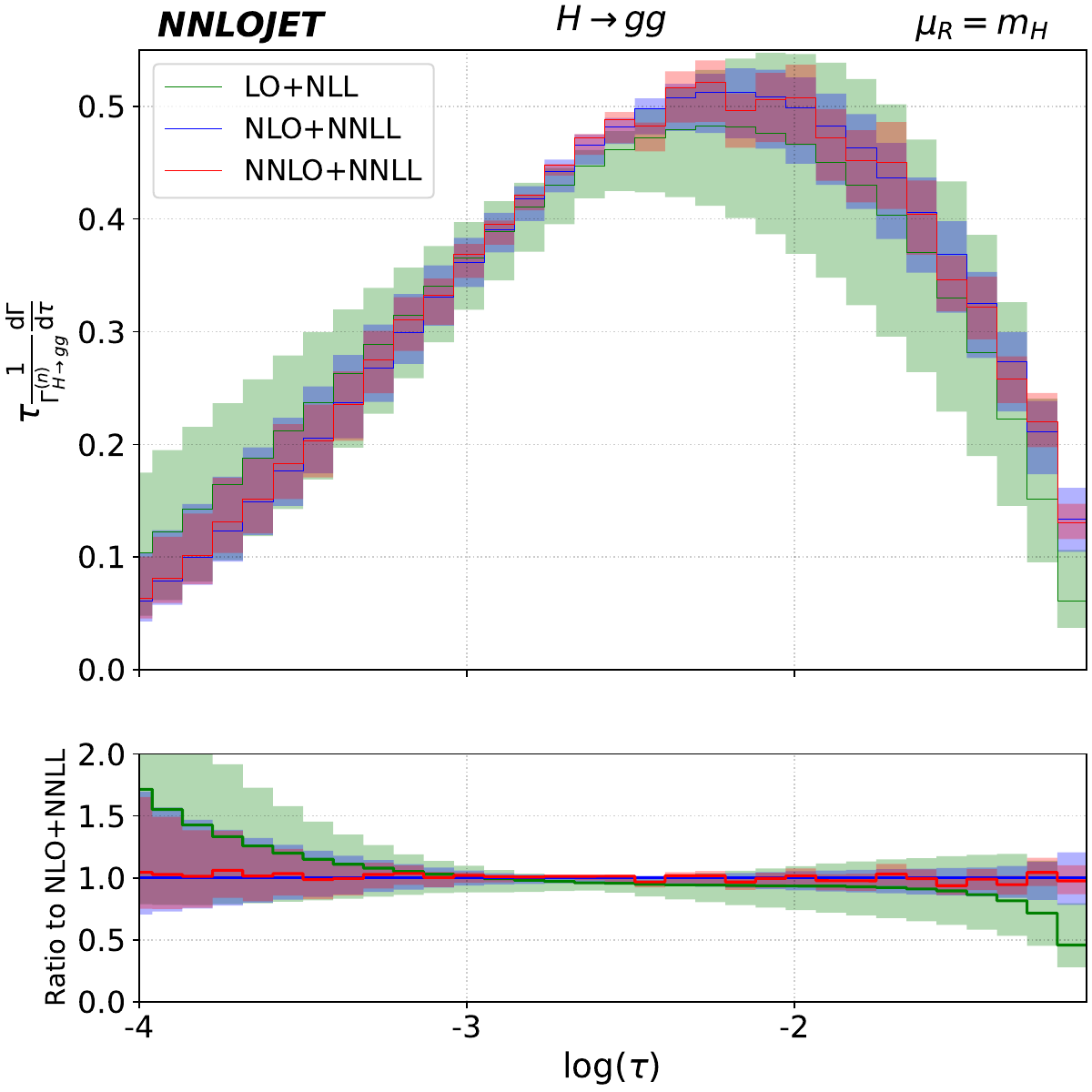}
  \caption{Thrust distribution results for fixed-order predictions matched to resummation in the logR scheme, for the $H\to q\bar{q}$ channel (left column) and the $H\to gg$ channel (right column), on a linear scale (top row) and on a log scale (bottom row). The curves represent the LO calculation matched to NLL resummation (green), NLO matched to NNLL (blue) and NNLO matched to NNLL (red).}
  \label{fig:matched}
\end{figure}

In Fig.~\ref{fig:matched}, we show matched predictions of the thrust observable in $H\to q\bar{q}$ and $H\to gg$ decays at LO+NLL, NLO+NNLL, and NNLO+NNLL.
The first row shows results on a linear $x$-axis, while in the bottom row we present results on a logarithmic axis, to better expose the resummation region.
The purpose of this figure is to show the progression of the predictions from LO to NNLO, matched to the resummation.
Except for the NNLO result, the fixed-order predictions are matched to the logarithmic order at which the logarithmic structure of the fixed-order calculation is fully captured for the first time, i.e. NLL for LO and NNLL for NLO.
The $x$-axis of the figures is limited to the region below the kinematical endpoint of the three-particle configuration, $\tau = 1/3$, where large logarithmic corrections due to Sudakov shoulder effects~\cite{Catani:1997xc} set in.
In both decay channels, the NLO+NNLL prediction leads to sizeable corrections compared to the baseline LO+NLL prediction, with further visible changes above $\tau = 0.2$ upon including the NNLO correction, as can be inferred from the linear plots.
At the three-particle kinematical endpoint, $\tau = 1/3$, the correction is about 40\% in the $H\to q\bar{q}$ and about 20\% in the $H\to gg$ channel.
The size of the uncertainty band is visibly reduced at NNLO in the hard region towards the right-hand side of the plots.
The logarithmic plots in the bottom row reveal that the reduced scale uncertainty at NNLO+NNLL compared to the NLO+NNLL carries over also to the region of intermediate $\tau$, but then becomes comparable between the two predictions below $\log(\tau) \approx -4.5$ in the $H\to q\bar{q}$ and $\log(\tau)\approx -3.0$ in the $H\to gg$ channel, where resummation effects dominate.

\begin{figure}[t]
  \centering
  \includegraphics[width=0.48\textwidth]{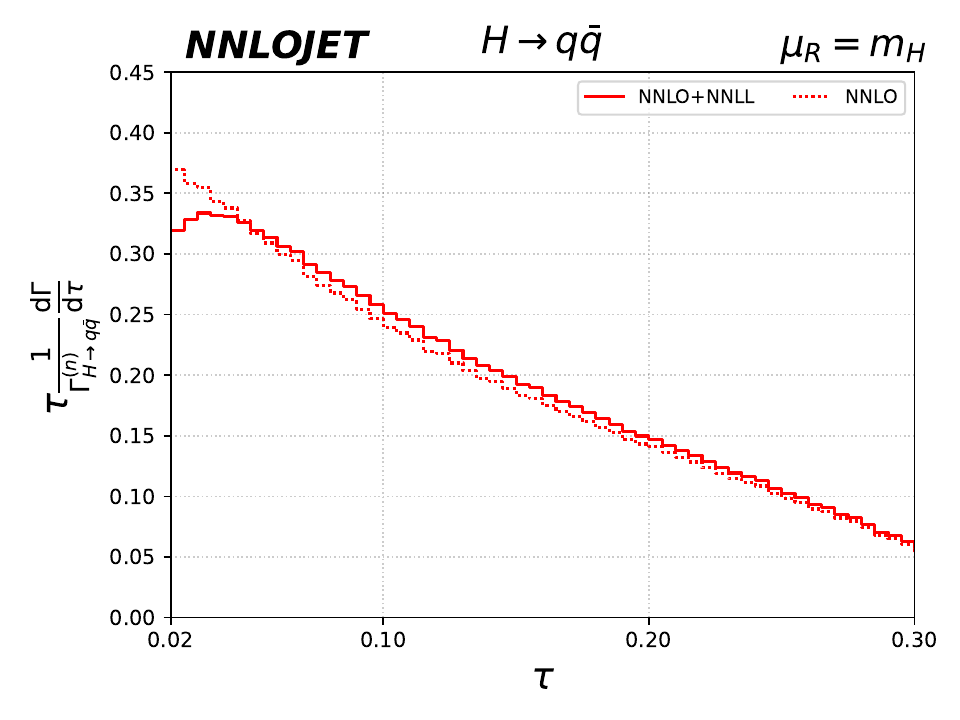}
  \includegraphics[width=0.48\textwidth]{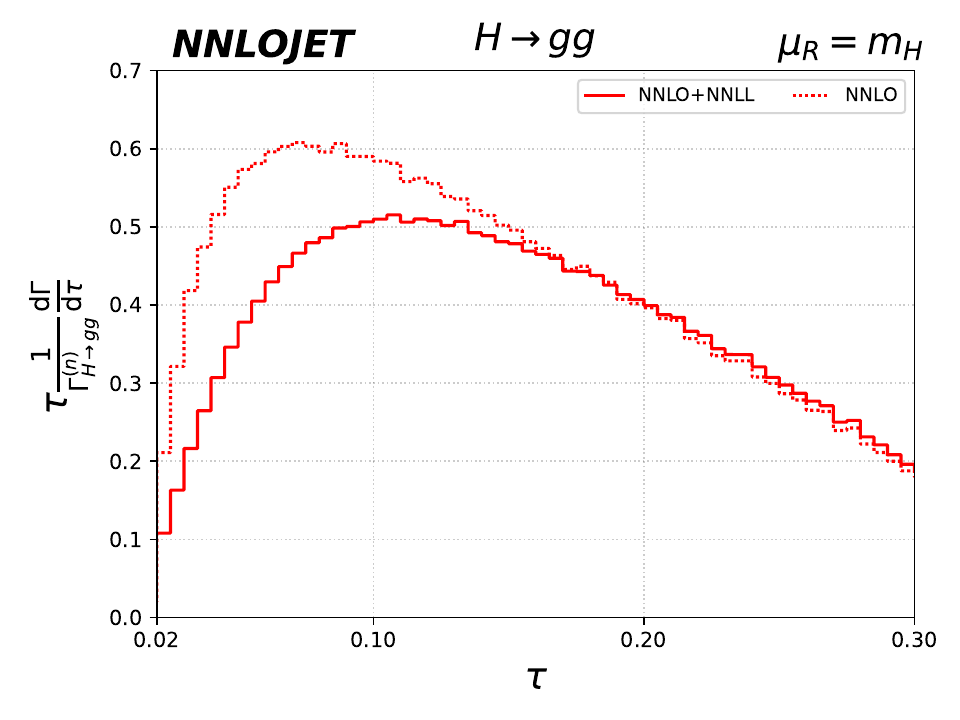}
  \caption{Comparison between NNLO (dashed line) and NNLO+NNLL (solid line) predictions for the thrust distribution in the hadronic decay of a Higgs boson in the Yukawa (left) and gluonic (right) modes.}
  \label{fig:matched_vs_FO}
\end{figure}

Fig.~\ref{fig:matched_vs_FO} contains a comparison of the fixed-order calculation at NNLO to the matched prediction at NNLO+NNLL. 
The left-hand plot shows the comparison in the $H\to q\bar{q}$ decay mode, while the right-hand plot shows the same comparison for the $H\to gg$ decay mode.
In both cases, the matched NNLO+NNLL prediction approaches the pure NNLO prediction in the limit $\tau \to 1/3$, but leads to visible differences when moving to the left in the plots, $\tau \to 0$.
For the $H\to q\bar{q}$ decay, the matched prediction only differs from the fixed-order result at the very left edge of the plot, at around $\tau \approx 0.05$, whereas in the $H\to gg$ mode, the matched and fixed-order predictions visibly differ already at around $\tau \approx 0.15$.
As expected, the peak of the distribution is shifted away from the infrared region in both Higgs decay channels when matched predictions are considered.

\begin{figure}[t]
  \centering
  \includegraphics[width=0.48\textwidth]{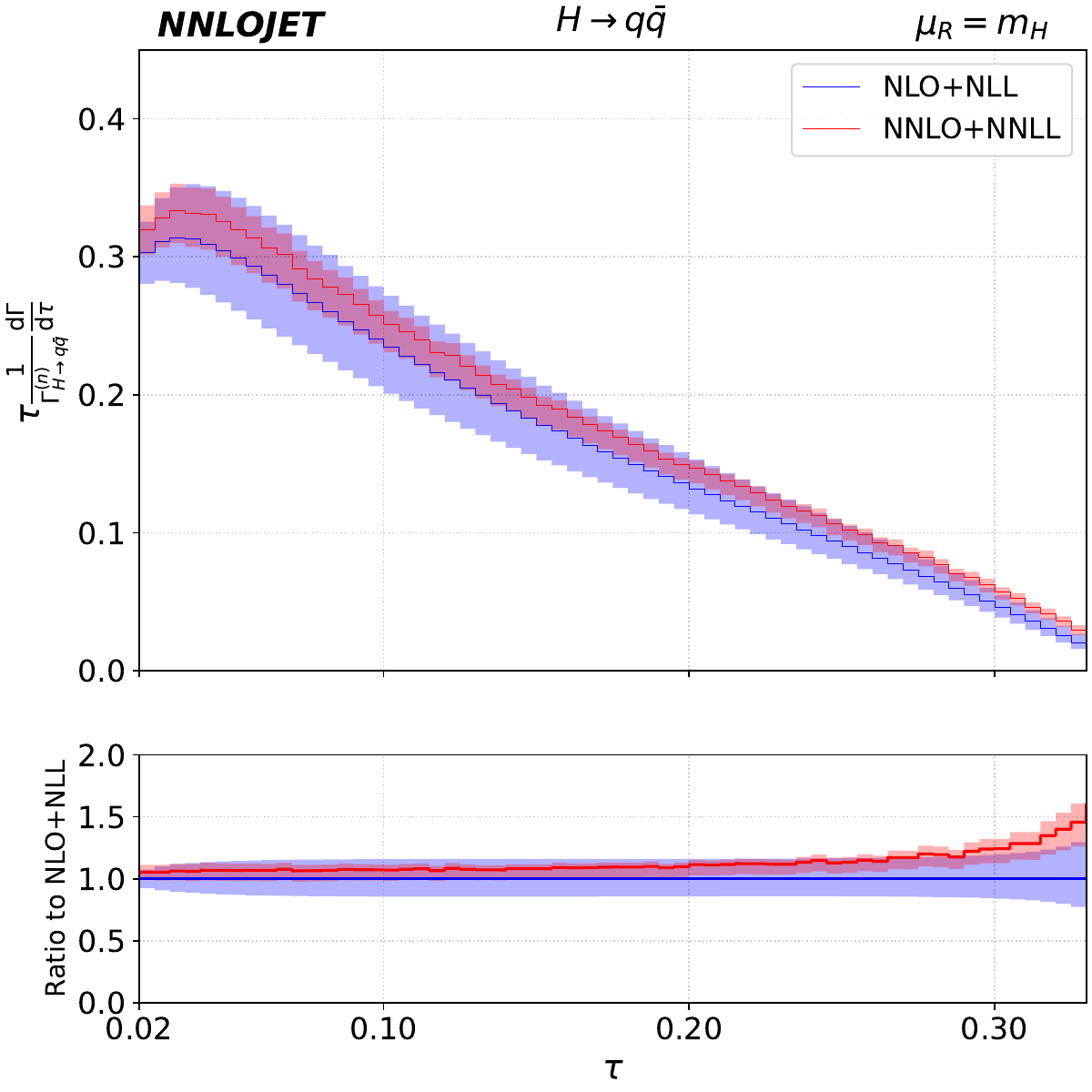}
  \includegraphics[width=0.48\textwidth]{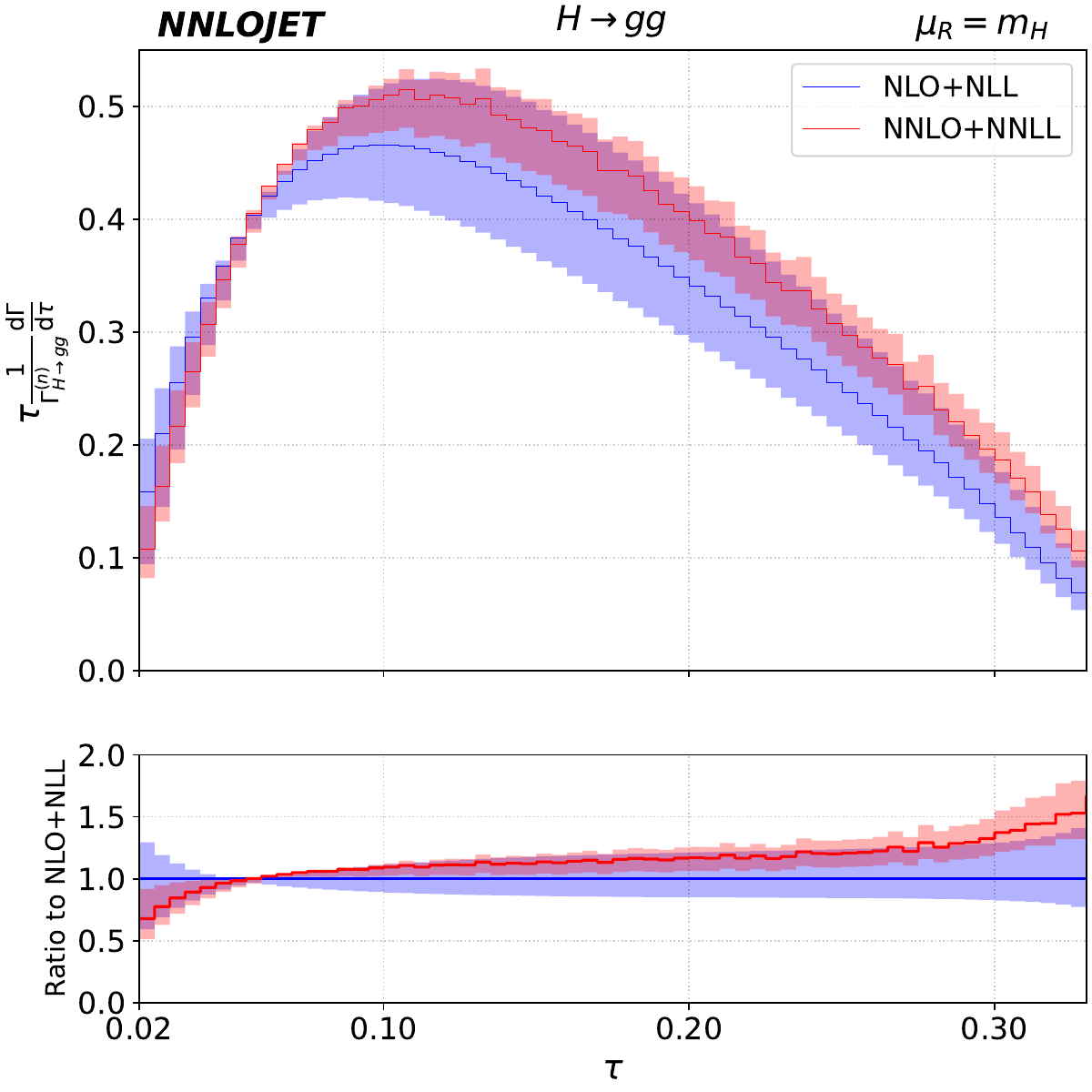}
  \includegraphics[width=0.48\textwidth]{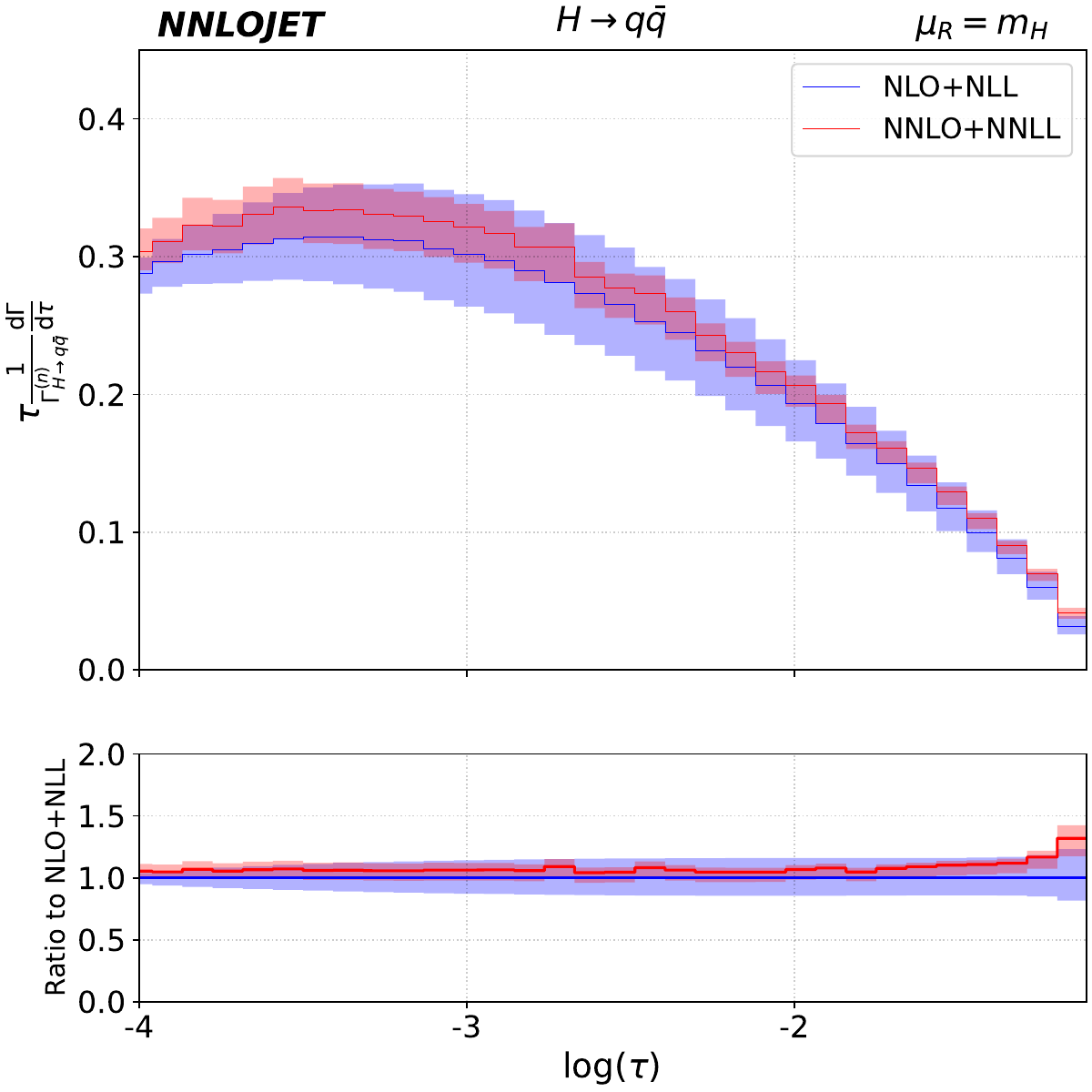}
  \includegraphics[width=0.48\textwidth]{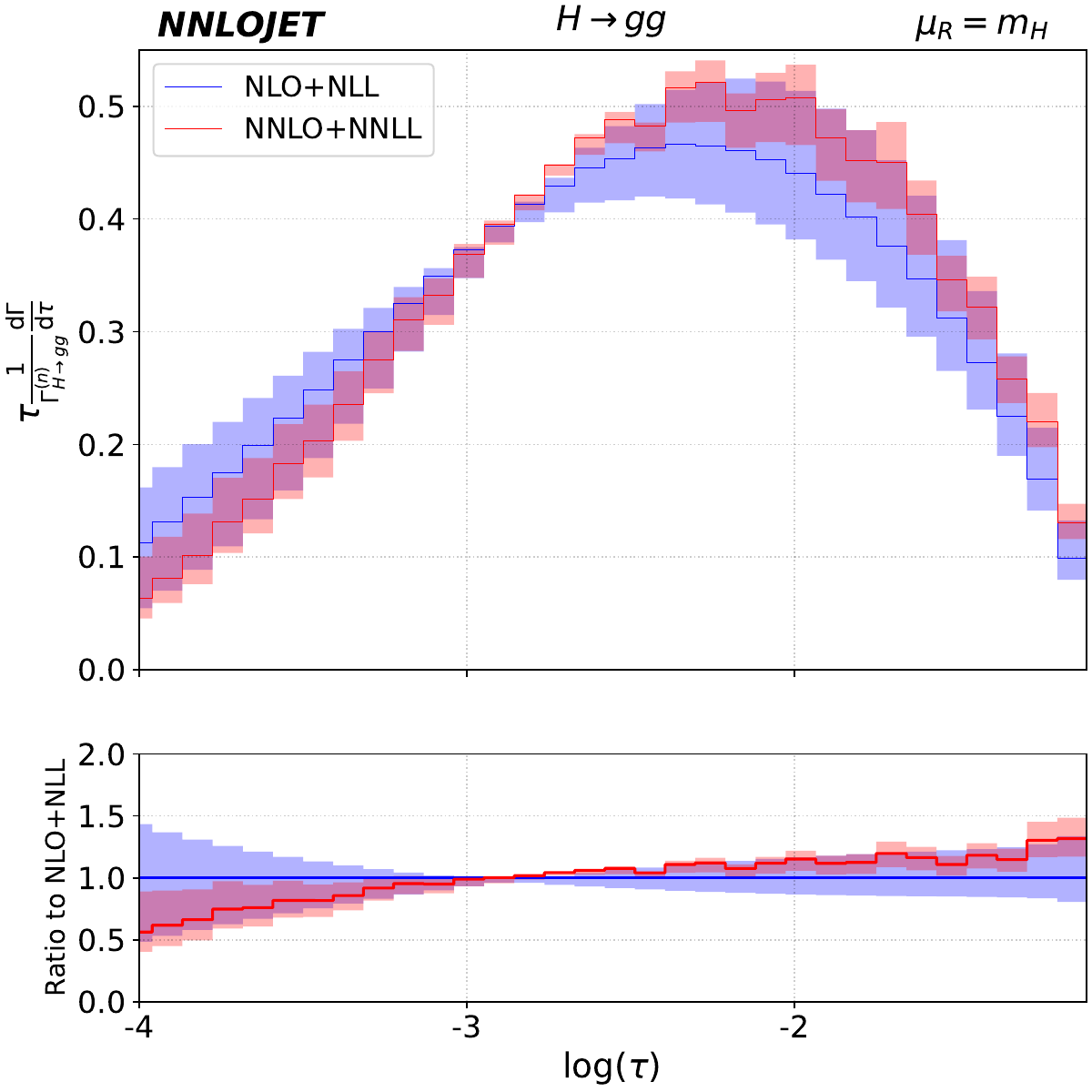}
  \caption{Comparison between NLO+NLL (blue) and NNLO+NNLL (red) results for the thrust distribution in the hadronic decay of a Higgs boson in the Yukawa (left) and gluonic (right) decay modes.}
  \label{fig:nn_vs_n}
\end{figure}

We compare our new NNLO+NNLL results to the previously achieved NLO+NLL accuracy \cite{Gehrmann-DeRidder:2024avt} in Fig.~\ref{fig:nn_vs_n}, where the upper row shows the comparison on a linear $x$-axis, while the lower row shows the same comparison on a logarithmic scale, emphasising the resummation region.
The range of the $x$-axis is limited from below by $\tau = 0.02$ and logarithmic plots by $\log(\tau) = -4.0$ in both Higgs decay modes.
In both cases, the upper limit is taken to be the three-particle kinematical boundary, $\tau=1/3$, where large logarithms from Sudakov shoulder effects arise.
The purpose of this comparison is to highlight the improvements obtained with our calculation in comparison to the previous prediction of \cite{Gehrmann-DeRidder:2024avt}.
As can be inferred from both sets of plots, the NNLO+NNLL correction is positive over most of the domain, with only a small region in the deep infrared which receives negative contributions.
This observation is in line with both the pure NNLO and pure NNLL corrections with respect to the NLO and NLL predictions, respectively, see Fig.~\ref{fig:resummed} above and Fig.~6 in \cite{Fox:2025qmp}.
In both decay modes, the NNLO+NNLL correction is sizeable over the entire spectrum, reflecting the inclusion of higher orders in both the fixed-order and the resummed calculation.
To the right of the plot, around the three-particle kinematical limit, a positive correction of about 50\% can be seen in both decay modes.
In the $H\to gg$ decay, a large negative correction of around 50\% is also noted towards the left of the plots, in the infrared region.
In both decay modes, the scale uncertainties are visibly reduced, thanks to the NNLO correction at large values of $\tau$ and the NNLL correction in the infrared region, $\tau \to 0$.
This observation holds also in the intermediate region.
While the peak remains roughly in the same position for $H\to q\bar{q}$ upon including the NNLO+NNLL corrections, it visibly shifts to the right for $H\to gg$.
Except for the small region very close to $\tau = 1/3$ in the $H\to b\bar{b}$ decay mode, the uncertainty bands of the NLO+NLL calculation overlap with the ones from the NNLO+NNLL calculation over the entire range in both decay modes, indicating very good convergence of the perturbative series for thrust.

\begin{figure}[t]
  \centering
  \includegraphics[width=0.48\textwidth]{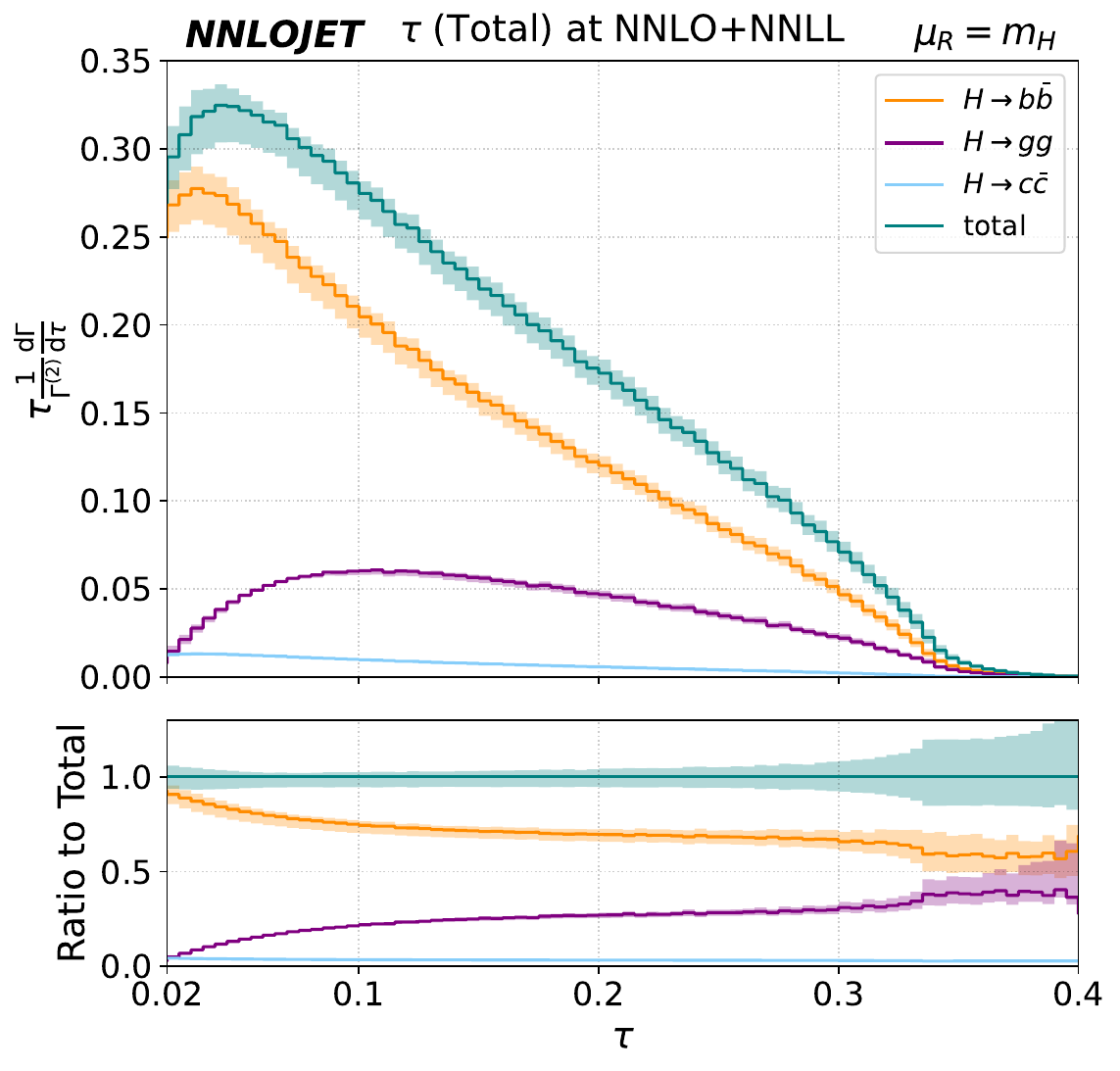}
  \includegraphics[width=0.48\textwidth]{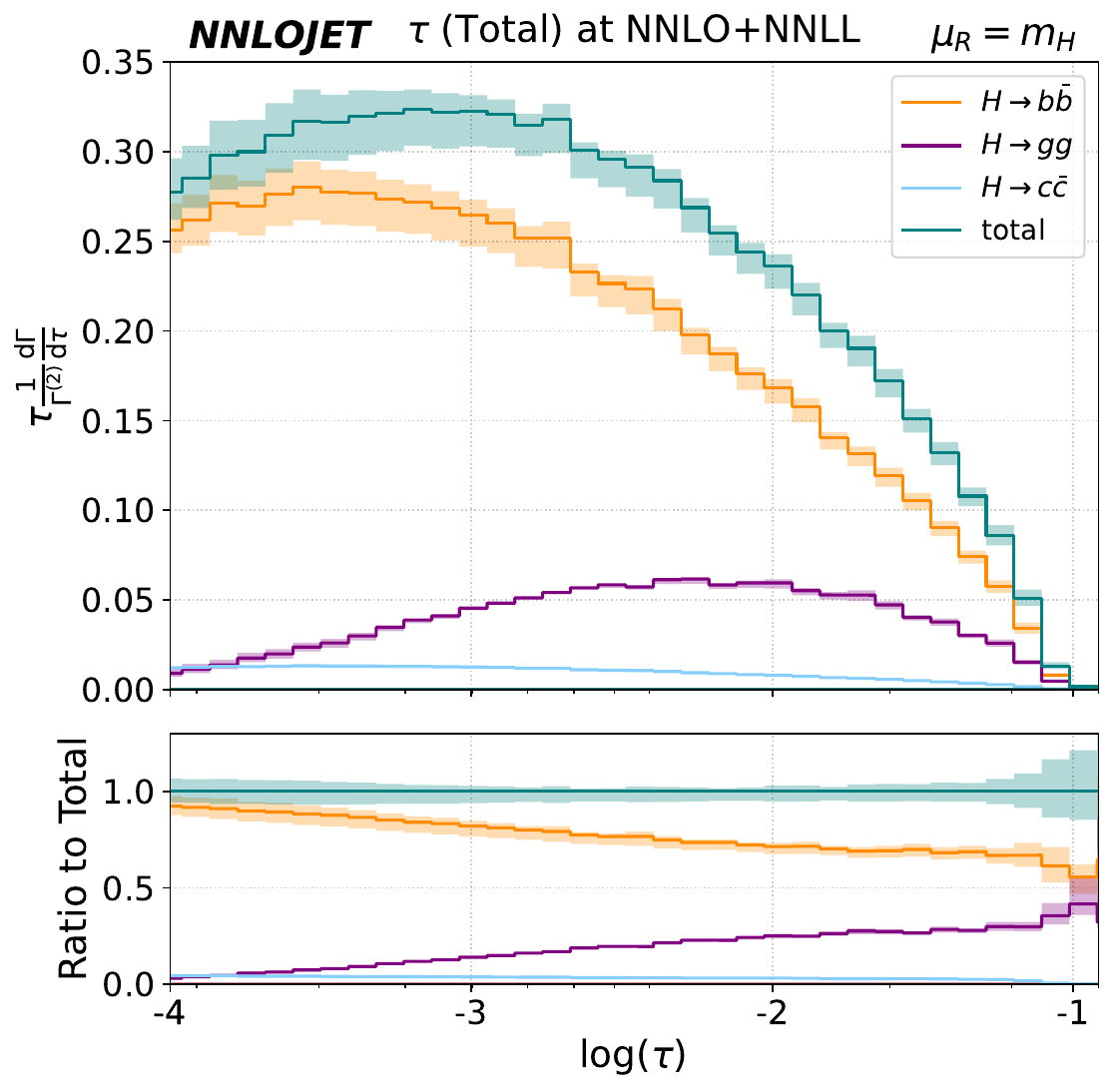}
  \caption{Results for the thrust distribution in the hadronic decay of a Higgs boson for a NNLO fixed-order calculation matched in the logR scheme to NNLL resummation. Curves for the total sum over all decay channels (teal), the $H\to b\bar{b}$ channel (orange), the $H\to gg$ channel (purple) and the $H\to c\bar{c}$ (light blue) are presented. The ratio to the total sum is shown in the lower frames.} 
  \label{fig:matched_total}
\end{figure}

Finally, we show matched predictions for the sum over the $H\to b\bar{b}$, $H\to c\bar{c}$, and $H\to gg$ channels in Fig.~\ref{fig:matched_total}.
Here, the left-hand side plot shows the thrust distributions at NNLO+NNLL on a linear $x$-axis, while the right-hand side plot shows the same distributions on a logarithmic $x$-axis.
The linear scale is limited from above by the five-particle kinematical endpoint, while the logarithmic axis is limited from below by the $\log(\tau)=-4.0$.
Generally, the plots confirm the findings of \cite{Fox:2025qmp}.
Specifically, we observe the same enhancement of $H\to gg$ for large $\tau$ as observed in the pure fixed-order case, which is expected, because we correctly recover the NNLO calculation in this limit.
Towards the five-particle kinematical limit, the $H\to b\bar{b}$ and $H\to gg$ contributions approach each other, with the former yielding a contribution slightly above 50\% and the latter yielding a contribution of slightly less than 50\% at the endpoint towards the right of the plot.
As can be observed from the logarithmic plots, the $H\to gg$ fraction decreases as $\tau\to 0$, even below the $H\to c\bar{c}$ contribution, confirming the trend observed in the pure NNLO study \cite{Fox:2025qmp}, which however did not faithfully describe this region. Nevertheless, it should be noted, that hadronisation as well as quark mass effects become relevant in this region of phase space.
In general, even in the intermediate $\tau$ region and around the peak of the $H\to gg$ contribution, the NNLL matching does not change the relative fraction of $H\to gg$ and $H\to b\bar{b}$ much with respect to the pure NNLO calculation.

\section{Conclusions}
\label{sec:conclusion}
Hadronic  final states in Higgs boson decays will be explored at future $e^+e^-$ colliders, offering unique 
opportunities for probing the various hadronic decay channels of the Higgs boson. To enable precision 
QCD studies in these final states, 
we have computed predictions for the thrust observable in hadronic Higgs decays to quarks and gluons at NNLO+NNLL accuracy.
The fixed-order calculation up to NNLO was performed using the \NNLOJET parton-level event generator, using the antenna subtraction scheme.
The resummation up to NNLL was carried out in a stand-alone implementation of the \ARES scheme.

Compared to the baseline fixed-order NNLO calculation presented in \cite{Fox:2025qmp}, the matched NNLO+NNLL prediction exhibits a visible shift of the peak of the thrust distribution away from the infrared region in both decay modes. As expected, this is more pronounced in the $H\to gg$ decay mode than in the $H\to q\bar{q}$ channel.
The NNLO+NNLL corrections are substantial relative to the NLO+NLL prediction \cite{Gehrmann-DeRidder:2024avt}, reaching up to 50\% at the edges of the spectrum.
Most notably, the NNLO+NNLL matching leads to significantly reduced scale-uncertainty bands compared to both the NLO+NLL and NLO+NNLL calculations.

Besides their importance for precision Higgs phenomenology, these predictions serve as valuable benchmarks for parton-shower algorithms and matching schemes incorporating higher logarithmic accuracy \cite{Dasgupta:2020fwr,Forshaw:2020wrq,Herren:2022jej,Assi:2023rbu,Hoche:2024dee,Hamilton:2023dwb,FerrarioRavasio:2023kyg,Preuss:2024vyu,vanBeekveld:2025lpz,Hoche:2025gsb}.
Given the central role of the thrust observable in both experimental analyses and theoretical studies and considering its relatively simple structure, an extension to NNLO+\N{3}LL appears feasible.
At this logarithmic order, full control over terms of the form $\alphas^nL^{n-2}$ will be achieved, corresponding to the single-logarithmic $\alphas^3L$ contributions at NNLO.
Since both our fixed-order calculation and the implementation of the \ARES resummation scheme are, in principle, general, we plan to extend this study to other event-shape observables in future work.

\acknowledgments
The authors would like to thank Basem El-Menoufi for discussions on the \ARES scheme.
AG acknowledges the support of the Swiss National Science Foundation (SNF) under contract 200021-231259 and of the Swiss National Supercomputing Centre (CSCS) under project ID ETH5f. 
TG has received funding from the Swiss National Science Foundation (SNF) under contract 10005816 and from the European Research Council (ERC) under the European Union's Horizon 2020 research and innovation programme grant agreement 101019620 (ERC Advanced Grant TOPUP).
NG gratefully acknowledges support from the UK Science and Technology Facilities Council (STFC) under contract ST/X000745/1 and hospitality from the Pauli Center for Theoretical Studies, Zurich. 
Part of the computations were carried out on the PLEIADES cluster at the University of Wuppertal, supported by the Deutsche Forschungsgemeinschaft (DFG, grant No. INST 218/78-1 FUGG) and the Bundesministerium f{\"u}r Bildung und Forschung (BMBF).
MM is supported by a Royal Society Newton International Fellowship (NIF/R1/232539).

\appendix
\section{Inclusive Hadronic Decay Rates}\label{app:A}
The Higgs-boson inclusive  decay rates for the two modes we consider read
\begin{align}
  \Gamma^{(k)}_{H\to q\bar{q}} &= \Gamma^{(0)}_{H\to q\bar{q}}\, \left(1+\sum\limits_{n=1}^k\left(\dfrac{\alphas}{2\pi}\right)^n H_{q\bar{q}}^{(n)}\right) \,,\\
  \Gamma^{(k)}_{H\to gg} &= \Gamma^{(0)}_{H\to gg}\, \left(1+\sum\limits_{n=1}^k\left(\dfrac{\alphas}{2\pi}\right)^n H_{gg}^{(n)}\right) \,.
  \label{eq:ratesNkLO1}
\end{align}
The perturbative corrections~\cite{Herzog:2017dtz}  are given at NLO by
\begin{align}
  H_{q\bar{q}}^{(1)} &=  \dfrac{r_1}{2}+2\gamma_0 L_R\,,\\
  H_{gg}^{(1)} &= c_1+\dfrac{1}{2}g_1+4\pi\beta_0 L_R\,,
\end{align}
and at NNLO by
\begin{align}
  H_{q\bar{q}}^{(2)} &=\dfrac{1}{4}\left(r_2+(8\gamma_1+4r_1\gamma_0+4\pi r_1\beta_0)L_R+(8\gamma_0^2+8\pi\beta_0\gamma_0)L_R^2\right)\,,\\
  H_{gg}^{(2)} &=  \dfrac{1}{4}\left(2d_2+d_1^2+g_2+(64\pi^2\beta_1+12\pi\beta_0g_1)L_R+48\pi^2\beta_0^2L_R^2+2d_1(g_1+8\pi\beta_0L_R)\right)\,,
\end{align}
where
\begin{equation}
  L_R = \log\frac{\mu_R^2}{m_H^2}\,,
\end{equation}
and the numerical constants are
\begin{equation}
	\begin{split}
		r_1 &= 17\CF \,,\\
		g_1 &= \dfrac{73}{3}\CA-\dfrac{14}{3}\NF \,,\\
		d_1 &= 11\,,\\
		r_2 &= \CF^2\left(\dfrac{691}{4}-36\zeta_2-36\zeta_3\right)+\CA\CF\left(\dfrac{893}{4}-22\zeta_2-62\zeta_3\right)\,,\\
		& -\CF\NF\left(\dfrac{65}{2}-4\zeta_2-8\zeta_3\right)\,,\\
		g_2 &=  \CA^2\left(\dfrac{37631}{54}-\dfrac{242}{3}\zeta_2-110\zeta_3\right)-\CA\NF\left(\dfrac{6665}{27}-\dfrac{88}{3}\zeta_2+4\zeta_3\right)\,,\\
		& -\CF\NF\left(\dfrac{131}{3}-24\zeta_3\right)+\NF^2\left(\dfrac{508}{27}-\dfrac{8}{3}\zeta_2\right)\,,\\
		d_2 &= \dfrac{2777}{18}+19\log\left(\dfrac{\mu_R^2}{m_t^2}\right)-\NF\left(\dfrac{67}{6}-\dfrac{16}{3}\log\left(\dfrac{\mu_R^2}{m_t^2}\right)\right)\,.
	\end{split}
\end{equation}
The running of the strong coupling is defined by
\begin{equation}
  \begin{split}
    \dfrac{d\alpha_s}{d\log(\mu_R^2)} &=\mu_R^2\dfrac{d\alpha_s}{d\mu_R^2}=-\alpha_s\left(\beta_0\alpha_s+\beta_1\alpha_s^2+\beta_2\alpha_s^3+\mathcal{O}(\alpha_s^3)\right)\,,
  \end{split}
\end{equation}
with
\begin{equation}
  \begin{split}
    \beta_0 &= \dfrac{1}{2\pi}\left(\dfrac{11}{6}\CA-\dfrac{1}{3}\NF\right)\,,\\
    \beta_1 &= \dfrac{1}{(2\pi)^2}\left(\dfrac{17}{6}\CA^2-\dfrac{5}{6}\CA\NF-\dfrac{1}{2}\CF\NF\right)\,,\\
    \beta_2 &= \dfrac{1}{(2\pi)^3}\left(\dfrac{2857}{432}\CA^3+\dfrac{54\CF^2-615\CA\CF-1415\CA^2}{432}\NF+\dfrac{66\CF+79\CA}{432}\NF^2\right)\,.
  \end{split}
\end{equation}
The running of Yukawa coupling is defined by
\begin{equation}
	\begin{split}
		\dfrac{dy}{d\log(\mu_R^2)} &=\mu_R^2\dfrac{dy}{d\mu_R^2}=-y\left(\gamma_0\left(\dfrac{\alpha_s}{2\pi}\right)+\gamma_1\left(\dfrac{\alpha_s}{2\pi}\right)^2+\mathcal{O}(\alpha_s^3)\right)\,,
	\end{split}
\end{equation}
with
\begin{equation}
	\begin{split}
		\gamma_0 &= \dfrac{3}{2}\CF\,,\\
		\gamma_1 &= \dfrac{1}{4}\left(\dfrac{3}{2}\CF^2+\dfrac{97}{6}\CF\CA-\dfrac{5}{3}\CF\NF\right)\,.
	\end{split}
\end{equation}

\section{Renormalisation Scale Dependence}\label{app:B}
For the Yukawa channel, the renormalisation scale dependence of the expansion coefficients is given by
\begin{equation}
  \begin{split}
    A_{q\bar{q}}(\mu_R) &= A_{q\bar{q}}(\mu_0) \,, \\
    B_{q\bar{q}}(\mu_R) &= B_{q\bar{q}}(\mu_0)+(2\pi\beta_0+2\gamma_0)L_RA_{q\bar{q}}(\mu_0) \,, \\
    C_{q\bar{q}}(\mu_R) &= C_{q\bar{q}}(\mu_0)+(4\pi\beta_0+2\gamma_0)L_RB_{q\bar{q}}(\mu_0) \\
    & +\left((4\pi^2\beta_1+2\gamma_1)L_R+(2\pi\beta_0+\gamma_0)(2\pi\beta_0+2\gamma_0)L_R^2)\right)A_{q\bar{q}}(\mu_0) \,.
  \end{split}
\end{equation}
where
\begin{equation}
L _R= \log\frac{\mu_R^2}{\mu_0^2}.
\end{equation}
For the gluonic channel:
\begin{equation}
  \begin{split}
  A_{gg}(\mu_R) &= A_{gg}(\mu_0)\,,\\
  B_{gg}(\mu_R) &= B_{gg}(\mu_0)+6\pi\beta_0L_RA_{gg}(\mu_0)\,,\\
  C_{gg}(\mu_R) &= C_{gg}(\mu_0)+8\pi\beta_0L_RB_{gg}(\mu_0)\,,\\
  &+\left(4\pi^2\beta_1L_R+24\pi^2\beta_0^2L_R^2-(2\pi d_1\beta_0-8\pi^2\beta_1)L_t\right)A_{gg}(\mu_0)\,,
  \end{split}
\end{equation}
where
\begin{equation}
  L_t = \log\left(\dfrac{m_t(\mu_R)^2}{m_t(\mu_0)^2}\right).
\end{equation}
The analogous formulae for the expansion coefficients can be obtained by combining the above with the results in App.~\ref{app:A}.

\section{Fixed-Order Expansion Coefficients}\label{app:C}
We report here the values of the fixed-order expansion coefficients $G_{ij}$ of the NNLL predictions in \eqref{eq:logDepA}--\eqref{eq:logDepC} up to the third order in $\alphas$.

For the $q\bar{q}$ radiator pair, they coincide with the ones obtained in Ref.~\cite{Becher:2008cf} and read:
\allowdisplaybreaks
\begin{eqnarray}
	G_{12,q\bar{q}}&=&-2\CF\,,\\
	G_{23,q\bar{q}}&=&-\dfrac{11}{3}\CA\CF+\dfrac{2}{3}\NF\,,\\
	G_{34,q\bar{q}}&=&-\dfrac{847}{108}\CA^2\CF+\frac{77}{27}\CA\CF\NF-\dfrac{7}{27}\CF\NF^2\,,\\
	G_{11,q\bar{q}}&=&3\CF\,,\\
	G_{22,q\bar{q}}&=&-\dfrac{169}{36}\CA\CF+\dfrac{\pi^2}{3}\CA\CF-\dfrac{4\pi^2}{3}\CF^2+\dfrac{11}{18}\CF\NF\,,\\
	G_{33,q\bar{q}}&=&\left(-\dfrac{3197}{108}+\dfrac{11\pi^2}{9}\right)\CA^2\CF-\dfrac{22\pi^2}{3}\CA\CF^2+\left(\dfrac{512}{27}-\dfrac{2\pi^2}{9}\right)\CA\CF\NF\nonumber\\
	&+&\left(1+\dfrac{4\pi^2}{3}\right)\CF^2\NF-\dfrac{17}{27}\CA\NF^2+\dfrac{64\zeta_{3}}{3}\CF^3\,,\\
	G_{21,q\bar{q}}&=&\left( \dfrac{57}{4} - 6\zeta_3 \right) \CA\CF
	+ \left( \dfrac{3}{4} + \pi^{2} - 4\zeta_3 \right) \CF^{2}
	+  -\dfrac{5}{2}\CF\NF\,,\\
	G_{32,q\bar{q}}&=&\left( -\dfrac{11323}{648} + \dfrac{85}{18}\pi^{2} - \dfrac{11}{90}\pi^{4} + 11\zeta_3 \right) \CA^{2}\CF + \left( \dfrac{11}{8} - \dfrac{70}{27}\pi^{2} + \dfrac{4}{9}\pi^{4} - 110\zeta_3 \right) \CA\CF^{2} \nonumber\\
	&+& \left( \dfrac{673}{324} - \dfrac{32}{27}\pi^{2} + 2\zeta_3 \right) \CA\CF\NF + \left( \dfrac{43}{12} + \dfrac{4}{27}\pi^{2} + 16\zeta_3 \right) \CF^{2}\NF \nonumber\\
	&+& \left( \dfrac{35}{162} + \dfrac{2}{27}\pi^{2} \right) \CF\NF^{2} + \left( \dfrac{8}{45}\pi^{4} - 48\zeta_3 \right) \CF^{3}\,.
\end{eqnarray}
For a $gg$ radiator pair, they read:
\begin{eqnarray}
	G_{12,gg} &=& -2\CA\,, \\
	G_{23,gg} &=& -\dfrac{11}{3}\CA^{2} + \dfrac{2}{3}\CA\NF\,, \\
	G_{34,gg} &=& -\dfrac{847}{108}\CA^{3} + \dfrac{77}{27}\CA^{2}\NF - \dfrac{7}{27}\CA\NF^{2} \,, \\
	G_{11,gg} &=& \dfrac{11}{3}\CA - \dfrac{2}{3}\NF\,, \\
	G_{22,gg} &=& -\left( \dfrac{49}{12} + \pi^{2} \right)\CA^{2}- \dfrac{1}{9}\CA\NF + \dfrac{1}{9}\NF^{2}\,, \\
	G_{33,gg} &=& -\left( \dfrac{9349}{324} + \dfrac{55}{9}\pi^{2} - \dfrac{64}{3}\zeta_3 \right)\CA^{3} + \left( \dfrac{457}{54} + \dfrac{10}{9}\pi^{2} \right)\CA^{2}\NF + \CA\CF\NF \nonumber\\
	&-& \CA\NF^{2} - \dfrac{2}{81}\NF^{3}\,, \\
	G_{21,gg} &=& \left( \dfrac{1025}{54} + \dfrac{11}{9}\pi^{2} - 10\zeta_3 \right)\CA^{2} - \left( \dfrac{158}{27} + \dfrac{2}{9}\pi^{2} \right)\CA\NF - \CF\NF +\dfrac{10}{27}\NF^{2}\,, \\
	G_{32,gg} &=& -\left( \dfrac{2545}{324} - \dfrac{203}{54}\pi^{2} + \dfrac{1}{2}\pi^{4} - \dfrac{473}{3}\zeta_3 \right)\CA^{3}- \left( \dfrac{2225}{324} + \dfrac{80}{27}\pi^{2} - \dfrac{98}{3}\zeta_3 \right)\CA^{2}\NF \nonumber\\
	&+& \left( \dfrac{11}{6} - 4\zeta_3 \right)\CA\CF\NF + \left( \dfrac{371}{162} + \dfrac{10}{27}\pi^{2} \right)\CA\NF^{2}+ \dfrac{1}{2}\CF\NF^{2} - \dfrac{10}{81}\NF^{3}\,.
\end{eqnarray}
The N$^3$LL coefficients $G_{31} $ are given in \eqref{eq:G31qqb} and \eqref{eq:G31gg}.

\bibliography{bibliography.bib}

\providecommand{\href}[2]{#2}\begingroup\raggedright\begin{thebibliography}{10}

\bibitem{Brandt:1964sa}
S.~Brandt, C.~Peyrou, R.~Sosnowski, and A.~Wroblewski, {\it {The Principal axis
  of jets. An Attempt to analyze high-energy collisions as two-body
  processes}},  {\em Phys. Lett.} {\bf 12} (1964) 57--61.

\bibitem{Farhi:1977sg}
E.~Farhi, {\it {A QCD Test for Jets}},  {\em Phys. Rev. Lett.} {\bf 39} (1977)
  1587--1588.

\bibitem{Ellis:1980wv}
R.~K. Ellis, D.~A. Ross, and A.~E. Terrano, {\it {The Perturbative Calculation
  of Jet Structure in $e^+ e^-$ Annihilation}},  {\em Nucl. Phys. B} {\bf 178}
  (1981) 421--456.

\bibitem{Gehrmann-DeRidder:2007nzq}
A.~Gehrmann-De~Ridder, T.~Gehrmann, E.~W.~N. Glover, and G.~Heinrich, {\it
  {Second-order QCD corrections to the thrust distribution}},  {\em Phys. Rev.
  Lett.} {\bf 99} (2007) 132002, [\href{http://arxiv.org/abs/0707.1285}{{\tt
  arXiv:0707.1285}}].

\bibitem{Weinzierl:2009ms}
S.~Weinzierl, {\it {Event shapes and jet rates in electron-positron
  annihilation at NNLO}},  {\em JHEP} {\bf 06} (2009) 041,
  [\href{http://arxiv.org/abs/0904.1077}{{\tt arXiv:0904.1077}}].

\bibitem{DelDuca:2016csb}
V.~Del~Duca, C.~Duhr, A.~Kardos, G.~Somogyi, and Z.~Tr{\'o}cs{\'a}nyi, {\it
  {Three-Jet Production in Electron-Positron Collisions at
  Next-to-Next-to-Leading Order Accuracy}},  {\em Phys. Rev. Lett.} {\bf 117}
  (2016) 152004, [\href{http://arxiv.org/abs/1603.08927}{{\tt
  arXiv:1603.08927}}].

\bibitem{DelDuca:2016ily}
V.~Del~Duca, C.~Duhr, A.~Kardos, G.~Somogyi, Z.~Sz{\H{o}}r,
  Z.~Tr{\'o}cs{\'a}nyi, and Z.~Tulip{\'a}nt, {\it {Jet production in the
  CoLoRFulNNLO method: event shapes in electron-positron collisions}},  {\em
  Phys. Rev. D} {\bf 94} (2016) 074019,
  [\href{http://arxiv.org/abs/1606.03453}{{\tt arXiv:1606.03453}}].

\bibitem{Catani:1991kz}
S.~Catani, G.~Turnock, B.~R. Webber, and L.~Trentadue, {\it {Thrust
  distribution in $e^+ e^-$ annihilation}},  {\em Phys. Lett. B} {\bf 263}
  (1991) 491--497.

\bibitem{Monni:2011gb}
P.~F. Monni, T.~Gehrmann, and G.~Luisoni, {\it {Two-Loop Soft Corrections and
  Resummation of the Thrust Distribution in the Dijet Region}},  {\em JHEP}
  {\bf 08} (2011) 010, [\href{http://arxiv.org/abs/1105.4560}{{\tt
  arXiv:1105.4560}}].

\bibitem{Becher:2008cf}
T.~Becher and M.~D. Schwartz, {\it {A precise determination of $\alpha_s$ from
  LEP thrust data using effective field theory}},  {\em JHEP} {\bf 07} (2008)
  034, [\href{http://arxiv.org/abs/0803.0342}{{\tt arXiv:0803.0342}}].

\bibitem{Abbate:2010xh}
R.~Abbate, M.~Fickinger, A.~H. Hoang, V.~Mateu, and I.~W. Stewart, {\it {Thrust
  at $N^{3}LL$ with Power Corrections and a Precision Global Fit for
  $\alpha_{s}(m_Z)$}},  {\em Phys. Rev. D} {\bf 83} (2011) 074021,
  [\href{http://arxiv.org/abs/1006.3080}{{\tt arXiv:1006.3080}}].

\bibitem{Aglietti:2025jdj}
U.~G. Aglietti, G.~Ferrera, W.-L. Ju, and J.~Miao, {\it {Thrust Distribution in
  Electron-Positron Annihilation at Full
  Next-to-Next-to-Next-to-Leading-Logarithmic Accuracy Including
  Next-to-Next-to-Leading-Order Terms in QCD}},  {\em Phys. Rev. Lett.} {\bf
  134} (2025) 251904, [\href{http://arxiv.org/abs/2502.01570}{{\tt
  arXiv:2502.01570}}].

\bibitem{OPAL:2004wof}
{\bf OPAL} Collaboration, G.~Abbiendi et~al., {\it {Measurement of event shape
  distributions and moments in e+ e- ---{\ensuremath{>}} hadrons at 91-GeV -
  209-GeV and a determination of alpha(s)}},  {\em Eur. Phys. J. C} {\bf 40}
  (2005) 287--316, [\href{http://arxiv.org/abs/hep-ex/0503051}{{\tt
  hep-ex/0503051}}].

\bibitem{L3:2004cdh}
{\bf L3} Collaboration, P.~Achard et~al., {\it {Studies of hadronic event
  structure in $e^{+} e^{-}$ annihilation from 30-GeV to 209-GeV with the L3
  detector}},  {\em Phys. Rept.} {\bf 399} (2004) 71--174,
  [\href{http://arxiv.org/abs/hep-ex/0406049}{{\tt hep-ex/0406049}}].

\bibitem{Bethke:2009ehn}
{\bf JADE} Collaboration, S.~Bethke, S.~Kluth, C.~Pahl, and J.~Schieck, {\it
  {Determination of the Strong Coupling alpha(s) from hadronic Event Shapes
  with O(alpha**3(s)) and resummed QCD predictions using JADE Data}},  {\em
  Eur. Phys. J. C} {\bf 64} (2009) 351--360,
  [\href{http://arxiv.org/abs/0810.1389}{{\tt arXiv:0810.1389}}].

\bibitem{Dissertori:2009qa}
G.~Dissertori, A.~Gehrmann-De~Ridder, T.~Gehrmann, E.~W.~N. Glover,
  G.~Heinrich, and H.~Stenzel, {\it {Precise determination of the strong
  coupling constant at NNLO in QCD from the three-jet rate in
  electron--positron annihilation at LEP}},  {\em Phys. Rev. Lett.} {\bf 104}
  (2010) 072002, [\href{http://arxiv.org/abs/0910.4283}{{\tt
  arXiv:0910.4283}}].

\bibitem{Dissertori:2009ik}
G.~Dissertori, A.~Gehrmann-De~Ridder, T.~Gehrmann, E.~W.~N. Glover,
  G.~Heinrich, G.~Luisoni, and H.~Stenzel, {\it {Determination of the strong
  coupling constant using matched NNLO+NLLA predictions for hadronic event
  shapes in $e^+e^-$ annihilations}},  {\em JHEP} {\bf 08} (2009) 036,
  [\href{http://arxiv.org/abs/0906.3436}{{\tt arXiv:0906.3436}}].

\bibitem{OPAL:2011aa}
{\bf OPAL} Collaboration, G.~Abbiendi et~al., {\it {Determination of $\alpha_s$
  using OPAL hadronic event shapes at $\sqrt{s}=91$ - 209 GeV and resummed NNLO
  calculations}},  {\em Eur. Phys. J. C} {\bf 71} (2011) 1733,
  [\href{http://arxiv.org/abs/1101.1470}{{\tt arXiv:1101.1470}}].

\bibitem{Benitez:2024nav}
M.~A. Benitez, A.~H. Hoang, V.~Mateu, I.~W. Stewart, and G.~Vita, {\it {On
  determining {\ensuremath{\alpha}}$_{s}$(m$_{Z}$) from dijets in
  e$^{+}$e$^{-}$ thrust}},  {\em JHEP} {\bf 07} (2025) 249,
  [\href{http://arxiv.org/abs/2412.15164}{{\tt arXiv:2412.15164}}].

\bibitem{Farren-Colloty:2025amh}
C.~Farren-Colloty, J.~Helliwell, R.~Patel, G.~P. Salam, and S.~Zanoli, {\it
  {Anomalous scaling of linear power corrections}},
  \href{http://arxiv.org/abs/2507.18696}{{\tt arXiv:2507.18696}}.

\bibitem{Nason:2025qbx}
P.~Nason and G.~Zanderighi, {\it {Fits of {\ensuremath{\alpha}}$_{s}$ from
  event-shapes in the three-jet region: extension to all energies}},  {\em
  JHEP} {\bf 06} (2025) 200, [\href{http://arxiv.org/abs/2501.18173}{{\tt
  arXiv:2501.18173}}].

\bibitem{Webber:1994cp}
B.~R. Webber, {\it {Estimation of power corrections to hadronic event shapes}},
   {\em Phys. Lett. B} {\bf 339} (1994) 148--150,
  [\href{http://arxiv.org/abs/hep-ph/9408222}{{\tt hep-ph/9408222}}].

\bibitem{Dokshitzer:1995zt}
Y.~L. Dokshitzer and B.~R. Webber, {\it {Calculation of power corrections to
  hadronic event shapes}},  {\em Phys. Lett. B} {\bf 352} (1995) 451--455,
  [\href{http://arxiv.org/abs/hep-ph/9504219}{{\tt hep-ph/9504219}}].

\bibitem{Agarwal:2020uxi}
N.~Agarwal, A.~Mukhopadhyay, S.~Pal, and A.~Tripathi, {\it {Power corrections
  to event shapes using eikonal dressed gluon exponentiation}},  {\em JHEP}
  {\bf 03} (2021) 155, [\href{http://arxiv.org/abs/2012.06842}{{\tt
  arXiv:2012.06842}}].

\bibitem{Caola:2021kzt}
F.~Caola, S.~Ferrario~Ravasio, G.~Limatola, K.~Melnikov, and P.~Nason, {\it {On
  linear power corrections in certain collider observables}},  {\em JHEP} {\bf
  01} (2022) 093, [\href{http://arxiv.org/abs/2108.08897}{{\tt
  arXiv:2108.08897}}].

\bibitem{Bhattacharya:2022dtm}
A.~Bhattacharya, M.~D. Schwartz, and X.~Zhang, {\it {Sudakov shoulder
  resummation for thrust and heavy jet mass}},  {\em Phys. Rev. D} {\bf 106}
  (2022) 074011, [\href{http://arxiv.org/abs/2205.05702}{{\tt
  arXiv:2205.05702}}].

\bibitem{Caola:2022vea}
F.~Caola, S.~Ferrario~Ravasio, G.~Limatola, K.~Melnikov, P.~Nason, and M.~A.
  Ozcelik, {\it {Linear power corrections to e$^{+}$e$^{-}$ shape variables in
  the three-jet region}},  {\em JHEP} {\bf 12} (2022) 062,
  [\href{http://arxiv.org/abs/2204.02247}{{\tt arXiv:2204.02247}}].

\bibitem{Nason:2023asn}
P.~Nason and G.~Zanderighi, {\it {Fits of {\ensuremath{\alpha}}$_{s}$ using
  power corrections in the three-jet region}},  {\em JHEP} {\bf 06} (2023) 058,
  [\href{http://arxiv.org/abs/2301.03607}{{\tt arXiv:2301.03607}}].

\bibitem{Dasgupta:2024znl}
M.~Dasgupta and F.~Hounat, {\it {Exploring soft anomalous dimensions for 1/Q
  power corrections}},  {\em JHEP} {\bf 09} (2025) 060,
  [\href{http://arxiv.org/abs/2411.16867}{{\tt arXiv:2411.16867}}].

\bibitem{Hoang:2025uaa}
A.~H. Hoang, V.~Mateu, M.~D. Schwartz, and I.~W. Stewart, {\it {Precision
  e$^{+}$e$^{-}$ hemisphere masses in the dijet region with power
  corrections}},  {\em JHEP} {\bf 09} (2025) 092,
  [\href{http://arxiv.org/abs/2506.09130}{{\tt arXiv:2506.09130}}].

\bibitem{FCC:2018byv}
{\bf FCC} Collaboration, A.~Abada et~al., {\it {FCC Physics Opportunities}:
  {Future Circular Collider Conceptual Design Report Volume 1}},  {\em Eur.
  Phys. J. C} {\bf 79} (2019) 474.

\bibitem{FCC:2018evy}
{\bf FCC} Collaboration, A.~Abada et~al., {\it {FCC-ee: The Lepton Collider}:
  {Future Circular Collider Conceptual Design Report Volume 2}},  {\em Eur.
  Phys. J. ST} {\bf 228} (2019) 261--623.

\bibitem{CEPCStudyGroup:2018ghi}
{\bf CEPC Study Group} Collaboration, M.~Dong et~al., {\it {CEPC Conceptual
  Design Report: Volume 2 - Physics \& Detector}},
  \href{http://arxiv.org/abs/1811.10545}{{\tt arXiv:1811.10545}}.

\bibitem{ILC:2013jhg}
{\bf ILC} Collaboration, H.~Baer et~al., {\it {The International Linear
  Collider Technical Design Report - Volume 2: Physics}},
  \href{http://arxiv.org/abs/1306.6352}{{\tt arXiv:1306.6352}}.

\bibitem{Gao:2016jcm}
J.~Gao, {\it {Probing light-quark Yukawa couplings via hadronic event shapes at
  lepton colliders}},  {\em JHEP} {\bf 01} (2018) 038,
  [\href{http://arxiv.org/abs/1608.01746}{{\tt arXiv:1608.01746}}].

\bibitem{Gao:2019mlt}
J.~Gao, Y.~Gong, W.-L. Ju, and L.~L. Yang, {\it {Thrust distribution in Higgs
  decays at the next-to-leading order and beyond}},  {\em JHEP} {\bf 03} (2019)
  030, [\href{http://arxiv.org/abs/1901.02253}{{\tt arXiv:1901.02253}}].

\bibitem{Gao:2020vyx}
J.~Gao, V.~Shtabovenko, and T.-Z. Yang, {\it {Energy-energy correlation in
  hadronic Higgs decays: analytic results and phenomenology at NLO}},  {\em
  JHEP} {\bf 02} (2021) 210, [\href{http://arxiv.org/abs/2012.14188}{{\tt
  arXiv:2012.14188}}].

\bibitem{Knobbe:2023njd}
M.~Knobbe, F.~Krauss, D.~Reichelt, and S.~Schumann, {\it {Measuring hadronic
  Higgs boson branching ratios at future lepton colliders}},  {\em Eur. Phys.
  J. C} {\bf 84} (2024) 83, [\href{http://arxiv.org/abs/2306.03682}{{\tt
  arXiv:2306.03682}}].

\bibitem{Coloretti:2022jcl}
G.~Coloretti, A.~Gehrmann-De~Ridder, and C.~T. Preuss, {\it {QCD predictions
  for event-shape distributions in hadronic Higgs decays}},  {\em JHEP} {\bf
  06} (2022) 009, [\href{http://arxiv.org/abs/2202.07333}{{\tt
  arXiv:2202.07333}}].

\bibitem{Gehrmann-DeRidder:2023uld}
A.~Gehrmann-De~Ridder, C.~T. Preuss, and C.~Williams, {\it {Four-jet event
  shapes in hadronic Higgs decays}},  {\em JHEP} {\bf 03} (2024) 104,
  [\href{http://arxiv.org/abs/2310.09354}{{\tt arXiv:2310.09354}}].

\bibitem{Gehrmann-DeRidder:2024avt}
A.~Gehrmann-De~Ridder, C.~T. Preuss, D.~Reichelt, and S.~Schumann, {\it
  {NLO+NLL' accurate predictions for three-jet event shapes in hadronic Higgs
  decays}},  {\em JHEP} {\bf 07} (2024) 160,
  [\href{http://arxiv.org/abs/2403.06929}{{\tt arXiv:2403.06929}}].

\bibitem{Fox:2025cuz}
E.~Fox, A.~Gehrmann-De~Ridder, T.~Gehrmann, N.~Glover, M.~Marcoli, and C.~T.
  Preuss, {\it {Jet Rates in Higgs Boson Decay at Third Order in QCD}},  {\em
  Phys. Rev. Lett.} {\bf 134} (2025) 251905,
  [\href{http://arxiv.org/abs/2502.17333}{{\tt arXiv:2502.17333}}].

\bibitem{Fox:2025qmp}
E.~Fox, A.~Gehrmann-De~Ridder, T.~Gehrmann, N.~Glover, M.~Marcoli, and C.~T.
  Preuss, {\it {Precise Predictions for Event Shapes in Hadronic Higgs
  Decays}},  \href{http://arxiv.org/abs/2508.14282}{{\tt arXiv:2508.14282}}.

\bibitem{Ma:2024qoa}
X.~Ma, Z.~Wu, J.~Wu, Y.~Huang, G.~Li, M.~Ruan, F.~L. Alves, S.~Jin, and
  L.~Shao, {\it {Measurements of decay branching fractions of the Higgs boson
  to hadronic final states at the CEPC}},  {\em Chin. Phys. C} {\bf 49} (2025)
  053001, [\href{http://arxiv.org/abs/2410.04465}{{\tt arXiv:2410.04465}}].

\bibitem{Mondini:2019gid}
R.~Mondini, M.~Schiavi, and C.~Williams, {\it {N$^{3}$LO predictions for the
  decay of the Higgs boson to bottom quarks}},  {\em JHEP} {\bf 06} (2019) 079,
  [\href{http://arxiv.org/abs/1904.08960}{{\tt arXiv:1904.08960}}].

\bibitem{Mondini:2019vub}
R.~Mondini and C.~Williams, {\it {$ H\to b\overline{b}j $ at
  next-to-next-to-leading order accuracy}},  {\em JHEP} {\bf 06} (2019) 120,
  [\href{http://arxiv.org/abs/1904.08961}{{\tt arXiv:1904.08961}}].

\bibitem{NNLOJET:2025rno}
{\bf NNLOJET} Collaboration, A.~Huss et~al., {\it {NNLOJET: a parton-level
  event generator for jet cross sections at NNLO QCD accuracy}},
  \href{http://arxiv.org/abs/2503.22804}{{\tt arXiv:2503.22804}}.

\bibitem{Mo:2017gzp}
J.~Mo, F.~J. Tackmann, and W.~J. Waalewijn, {\it {A case study of quark-gluon
  discrimination at NNLL' in comparison to parton showers}},  {\em Eur. Phys.
  J. C} {\bf 77} (2017) 770, [\href{http://arxiv.org/abs/1708.00867}{{\tt
  arXiv:1708.00867}}].

\bibitem{Ju:2023dfa}
W.-L. Ju, Y.~Xu, L.~L. Yang, and B.~Zhou, {\it {Thrust distribution in Higgs
  decays up to the fifth logarithmic order}},  {\em Phys. Rev. D} {\bf 107}
  (2023) 114034, [\href{http://arxiv.org/abs/2301.04294}{{\tt
  arXiv:2301.04294}}].

\bibitem{Banfi:2014sua}
A.~Banfi, H.~McAslan, P.~F. Monni, and G.~Zanderighi, {\it {A general method
  for the resummation of event-shape distributions in $e^+e^-$ annihilation}},
  {\em JHEP} {\bf 05} (2015) 102, [\href{http://arxiv.org/abs/1412.2126}{{\tt
  arXiv:1412.2126}}].

\bibitem{Banfi:2018mcq}
A.~Banfi, B.~K. El-Menoufi, and P.~F. Monni, {\it {The Sudakov radiator for jet
  observables and the soft physical coupling}},  {\em JHEP} {\bf 01} (2019)
  083, [\href{http://arxiv.org/abs/1807.11487}{{\tt arXiv:1807.11487}}].

\bibitem{Wilczek:1977zn}
F.~Wilczek, {\it {Decays of Heavy Vector Mesons Into Higgs Particles}},  {\em
  Phys. Rev. Lett.} {\bf 39} (1977) 1304.

\bibitem{Shifman:1978zn}
M.~A. Shifman, A.~I. Vainshtein, and V.~I. Zakharov, {\it {Remarks on Higgs
  Boson Interactions with Nucleons}},  {\em Phys. Lett. B} {\bf 78} (1978)
  443--446.

\bibitem{Inami:1982xt}
T.~Inami, T.~Kubota, and Y.~Okada, {\it {Effective Gauge Theory and the Effect
  of Heavy Quarks in Higgs Boson Decays}},  {\em Z. Phys. C} {\bf 18} (1983)
  69--80.

\bibitem{Vermaseren:1997fq}
J.~A.~M. Vermaseren, S.~A. Larin, and T.~van Ritbergen, {\it {The four loop
  quark mass anomalous dimension and the invariant quark mass}},  {\em Phys.
  Lett. B} {\bf 405} (1997) 327--333,
  [\href{http://arxiv.org/abs/hep-ph/9703284}{{\tt hep-ph/9703284}}].

\bibitem{Gehrmann-DeRidder:2005btv}
A.~Gehrmann-De~Ridder, T.~Gehrmann, and E.~W.~N. Glover, {\it {Antenna
  subtraction at NNLO}},  {\em JHEP} {\bf 09} (2005) 056,
  [\href{http://arxiv.org/abs/hep-ph/0505111}{{\tt hep-ph/0505111}}].

\bibitem{Currie:2013vh}
J.~Currie, E.~W.~N. Glover, and S.~Wells, {\it {Infrared Structure at NNLO
  Using Antenna Subtraction}},  {\em JHEP} {\bf 04} (2013) 066,
  [\href{http://arxiv.org/abs/1301.4693}{{\tt arXiv:1301.4693}}].

\bibitem{Fox:2024bfp}
E.~Fox, N.~Glover, and M.~Marcoli, {\it {Generalised antenna functions for
  higher-order calculations}},  {\em JHEP} {\bf 12} (2024) 225,
  [\href{http://arxiv.org/abs/2410.12904}{{\tt arXiv:2410.12904}}].

\bibitem{Braun-White:2023sgd}
O.~Braun-White, N.~Glover, and C.~T. Preuss, {\it {A general algorithm to build
  real-radiation antenna functions for higher-order calculations}},  {\em JHEP}
  {\bf 06} (2023) 065, [\href{http://arxiv.org/abs/2302.12787}{{\tt
  arXiv:2302.12787}}].

\bibitem{Braun-White:2023zwd}
O.~Braun-White, N.~Glover, and C.~T. Preuss, {\it {A general algorithm to build
  mixed real and virtual antenna functions for higher-order calculations}},
  {\em JHEP} {\bf 11} (2023) 179, [\href{http://arxiv.org/abs/2307.14999}{{\tt
  arXiv:2307.14999}}].

\bibitem{Herzog:2017dtz}
F.~Herzog, B.~Ruijl, T.~Ueda, J.~A.~M. Vermaseren, and A.~Vogt, {\it {On Higgs
  decays to hadrons and the R-ratio at N$^{4}$LO}},  {\em JHEP} {\bf 08} (2017)
  113, [\href{http://arxiv.org/abs/1707.01044}{{\tt arXiv:1707.01044}}].

\bibitem{Banfi:2004yd}
A.~Banfi, G.~P. Salam, and G.~Zanderighi, {\it {Principles of general
  final-state resummation and automated implementation}},  {\em JHEP} {\bf 03}
  (2005) 073, [\href{http://arxiv.org/abs/hep-ph/0407286}{{\tt
  hep-ph/0407286}}].

\bibitem{Catani:1992ua}
S.~Catani, L.~Trentadue, G.~Turnock, and B.~R. Webber, {\it {Resummation of
  large logarithms in $e^+e^-$ event shape distributions}},  {\em Nucl. Phys.
  B} {\bf 407} (1993) 3--42.

\bibitem{Ellis:1996mzs}
R.~K. Ellis, W.~J. Stirling, and B.~R. Webber, {\em {QCD and collider
  physics}}, vol.~8.
\newblock Cambridge University Press, 2, 2011.

\bibitem{Catani:1990rr}
S.~Catani, B.~R. Webber, and G.~Marchesini, {\it {QCD coherent branching and
  semiinclusive processes at large x}},  {\em Nucl. Phys. B} {\bf 349} (1991)
  635--654.

\bibitem{Arpino:2019ozn}
L.~Arpino, A.~Banfi, and B.~K. El-Menoufi, {\it {Near-to-planar three-jet
  events at NNLL accuracy}},  {\em JHEP} {\bf 07} (2020) 171,
  [\href{http://arxiv.org/abs/1912.09341}{{\tt arXiv:1912.09341}}].

\bibitem{Jones:2003yv}
R.~W.~L. Jones, M.~Ford, G.~P. Salam, H.~Stenzel, and D.~Wicke, {\it
  {Theoretical uncertainties on $\alpha_s$ from event shape variables in $e^+
  e^-$ annihilations}},  {\em JHEP} {\bf 12} (2003) 007,
  [\href{http://arxiv.org/abs/hep-ph/0312016}{{\tt hep-ph/0312016}}].

\bibitem{Spira:1997dg}
M.~Spira, {\it {QCD effects in Higgs physics}},  {\em Fortsch. Phys.} {\bf 46}
  (1998) 203--284, [\href{http://arxiv.org/abs/hep-ph/9705337}{{\tt
  hep-ph/9705337}}].

\bibitem{Actis:2008ug}
S.~Actis, G.~Passarino, C.~Sturm, and S.~Uccirati, {\it {NLO Electroweak
  Corrections to Higgs Boson Production at Hadron Colliders}},  {\em Phys.
  Lett. B} {\bf 670} (2008) 12--17, [\href{http://arxiv.org/abs/0809.1301}{{\tt
  arXiv:0809.1301}}].

\bibitem{Catani:1997xc}
S.~Catani and B.~R. Webber, {\it {Infrared safe but infinite: Soft gluon
  divergences inside the physical region}},  {\em JHEP} {\bf 10} (1997) 005,
  [\href{http://arxiv.org/abs/hep-ph/9710333}{{\tt hep-ph/9710333}}].

\bibitem{Dasgupta:2020fwr}
M.~Dasgupta, F.~A. Dreyer, K.~Hamilton, P.~F. Monni, G.~P. Salam, and G.~Soyez,
  {\it {Parton showers beyond leading logarithmic accuracy}},  {\em Phys. Rev.
  Lett.} {\bf 125} (2020), no.~5 052002,
  [\href{http://arxiv.org/abs/2002.11114}{{\tt arXiv:2002.11114}}].

\bibitem{Forshaw:2020wrq}
J.~R. Forshaw, J.~Holguin, and S.~Pl\"atzer, {\it {Building a consistent parton
  shower}},  {\em JHEP} {\bf 09} (2020) 014,
  [\href{http://arxiv.org/abs/2003.06400}{{\tt arXiv:2003.06400}}].

\bibitem{Herren:2022jej}
F.~Herren, S.~H\"oche, F.~Krauss, D.~Reichelt, and M.~Sch{\"o}nherr, {\it {A
  new approach to color-coherent parton evolution}},  {\em JHEP} {\bf 10}
  (2023) 091, [\href{http://arxiv.org/abs/2208.06057}{{\tt arXiv:2208.06057}}].

\bibitem{Assi:2023rbu}
B.~Assi and S.~H\"oche, {\it {A new approach to QCD evolution in processes with
  massive partons}},  \href{http://arxiv.org/abs/2307.00728}{{\tt
  arXiv:2307.00728}}.

\bibitem{Hoche:2024dee}
S.~H{\"o}che, F.~Krauss, and D.~Reichelt, {\it {alaric parton shower for hadron
  colliders}},  {\em Phys. Rev. D} {\bf 111} (2025), no.~9 094032,
  [\href{http://arxiv.org/abs/2404.14360}{{\tt arXiv:2404.14360}}].

\bibitem{Hamilton:2023dwb}
K.~Hamilton, A.~Karlberg, G.~P. Salam, L.~Scyboz, and R.~Verheyen, {\it
  {Matching and event-shape NNDL accuracy in parton showers}},  {\em JHEP} {\bf
  03} (2023) 224, [\href{http://arxiv.org/abs/2301.09645}{{\tt
  arXiv:2301.09645}}]. [Erratum: JHEP 11, 060 (2023)].

\bibitem{FerrarioRavasio:2023kyg}
S.~Ferrario~Ravasio, K.~Hamilton, A.~Karlberg, G.~P. Salam, L.~Scyboz, and
  G.~Soyez, {\it {Parton Showering with Higher Logarithmic Accuracy for Soft
  Emissions}},  {\em Phys. Rev. Lett.} {\bf 131} (2023), no.~16 161906,
  [\href{http://arxiv.org/abs/2307.11142}{{\tt arXiv:2307.11142}}].

\bibitem{Preuss:2024vyu}
C.~T. Preuss, {\it {A partitioned dipole-antenna shower with improved
  transverse recoil}},  {\em JHEP} {\bf 07} (2024) 161,
  [\href{http://arxiv.org/abs/2403.19452}{{\tt arXiv:2403.19452}}].

\bibitem{vanBeekveld:2025lpz}
M.~van Beekveld, S.~Ferrario~Ravasio, J.~Helliwell, A.~Karlberg, G.~P. Salam,
  L.~Scyboz, A.~Soto-Ontoso, G.~Soyez, and S.~Zanoli, {\it
  {Logarithmically-accurate and positive-definite NLO shower matching}},  {\em
  JHEP} {\bf 10} (2025) 038, [\href{http://arxiv.org/abs/2504.05377}{{\tt
  arXiv:2504.05377}}].

\bibitem{Hoche:2025gsb}
S.~H{\"o}che, F.~Krauss, P.~Meinzinger, and D.~Reichelt, {\it {Recoil-Safe
  Subtraction, Matching and Merging in $e^+e^- \to$ hadrons}},
  \href{http://arxiv.org/abs/2507.22837}{{\tt arXiv:2507.22837}}.

\end{thebibliography}\endgroup

\end{document}